\documentclass[12pt,reqno]{article}

\usepackage[latin1]{inputenc}
\usepackage[tbtags]{amsmath}
\usepackage{amsfonts}
\usepackage{amssymb}
\usepackage{amsxtra}
\usepackage{ushort}
\usepackage{color}
\usepackage{hyperref}
\hypersetup{linktocpage=true}

\def\be {\begin{equation}}
\def\ee {\end{equation}}
\def\bs#1\es{\begin{split}#1\end{split}}
\def\ba#1\ea{\begin{align}#1\end{align}}
\def\bg#1\eg{\begin{gathered}#1\end{gathered}}
\def\bea{\begin{eqnarray}}
\def\eea{\end{eqnarray}}

\def\ie{{\it i.e.}\ }
\def\eg{{\it e.g.}\ }

\def\a{\alpha}
\def\b{\beta}
\def\c{\chi}
\def\d{\delta}
\def\e{\epsilon}

\def\f{\phi}
\def\vf{\varphi}

\def\g{\gamma}
\def\G{\Gamma}
\def\h{\eta}

\def\l{\lambda}
\def\L{\Lambda}
\def\m{\mu}
\def\n{\nu}
\def\o{\omega}
\def\O{\Omega}
\def\p{\psi}

\def\r{\rho}
\def\s{\sigma}

\def\t{\tau}
\def\x{\xi}
\def\z{\zeta}

\def\pr{\prime} 
 
\def\pa{\partial}
\def\na{\nabla}
\def\fr{\frac}
\def\sq{\sqrt}

\def\bls{\bigg [}
\def\brs{\bigg ]}

\def\cL{\mathcal{L}}

\def\id{\rlap 1\mkern4mu{\rm l}}
\def\nn{\nonumber}

\def\on{\text{\tiny{(1)}}}
\def\tw{\text{\tiny{(2)}}}
\def\th{\text{\tiny{(3)}}}

\def\wh{\hat}
\def\we{\wedge} 

\def\ct{\otimes}
\newcommand{\un}[1]{ { \ushort{ #1}  } }
\newcommand{\uh}[1]{ {\hat {\ushort{ #1 } } } } 
\def\til{\tilde}

\numberwithin{equation}{section}
\numberwithin{figure}{section}
\title{\vspace{-1cm}
\begin{flushright}\normalsize
\hfill{MIFPA-11-39}\\
\hfill{Imperial/TP/11/KSS/03}
\end{flushright}
\vspace{1.5cm}
Chiral Reductions in the Salam-Sezgin Model}
\date{\vspace{-2cm}}

\newcommand{\hoch}[1]{$\, ^{#1}$}

\newcommand{\auth}{\large
C.N. Pope\footnote{email: pope@physics.tamu.edu}\hoch{\star}, T.G. Pugh\footnote{email: thomas.pugh08@imperial.ac.uk}\hoch{\dagger} and K.S. Stelle\footnote{email: k.stelle@imperial.ac.uk}\hoch{\dagger}}

\newcommand{\tamphys}{\normalsize\it George and Cynthia Woods Mitchell  Institute
for Fundamental Physics and Astronomy,\\
Texas A\&M University, College Station, TX 77843, USA\\
and\\
DAMTP, Centre for Mathematical Sciences,\\
 Cambridge University, Wilberforce Road, Cambridge CB3 OWA, UK}

\newcommand{\imperial}{\normalsize\it The Blackett Laboratory,
Imperial College London,\\ Prince Consort
Road, London SW7 2BZ, UK}

\author{}

\begin{document}

\setcounter{page}{0}
\maketitle
\thispagestyle{empty}

\begin{center}

\auth
\medskip

${}^\star$\tamphys

${}^\dagger$\imperial

\end{center}

\begin{abstract}
\normalsize
Reductions from six to four spacetime dimensions are considered for a class of supergravity models based on the six-dimensional Salam-Sezgin model, which is a chiral theory with a gauged $U(1)_R$ R-symmetry and a positive scalar-field potential. Reduction on a sphere and monopole background of such models naturally yields four-dimensional theories without a cosmological constant. The question of chirality preservation in such a reduction has been a topic of debate. In this article, it is shown that the possibilities of dimensional reduction bifurcate into two separate consistent dimensional-reduction schemes. One of these retains the massless $SU(2)$ vector gauge triplet arising from the sphere's isometries, but it produces a non-chiral four-dimensional theory. The other consistent scheme sets to zero the $SU(2)$ gauge fields, but retains the gauged $U(1)_R$ from six dimensions and preserves chirality although the $U(1)_R$ is spontaneously broken. Extensions of the Salam-Sezgin model to include larger gauge symmetries produce genuinely chiral models with unbroken gauge symmetries.
\end{abstract}

\newpage

\noindent\rule\textwidth{.1pt}		
\tableofcontents
\vspace{20pt}
\noindent\rule\textwidth{.1pt}

\setcounter{page}{1}

\section{Introduction} 
\label{Introduction} 

Applications of superstring and supergravity theories exploiting various special features of the very rich supergravity model set have dominated phenomenological approaches to physics beyond the Standard Model over the past three decades. Gauge symmetry possibilities, no-scale features, opportunities for supersymmetry breaking and other  specific features distinguish models within the landscape of possibilities. A striking feature of many of these models is the way in which they can be obtained via the many pathways of dimensional reduction from more uniquely defined theories in higher dimensions. 

A powerful example of this is the way in which the various options for gauge symmetries emerge from specific choice of reduction manifold \cite{Duff:1986hr,Hull:1988jw}. According to the choice of manifold, upon which a corresponding group action may be defined, one may obtain dimensionally reduced theories with compact or noncompact gaugings -- the latter arising from reduction on noncompact manifolds. This approach has been used, for example, to obtain a seven-dimensional model with gauged $SO(2,2)$ symmetry which, after further truncation and reduction to six dimensions, yields the chiral Salam-Sezgin \cite{Salam:1984cj} model with a gauged $U(1)_R$ R-symmetry \cite{Cvetic:2003xr}. The Salam-Sezgin model has in turn been taken as a key example of a theory that allows further reduction down to a four-dimensional theory without a cosmological constant even after supersymmetry breaking, thus providing a candidate solution to the cosmological constant problem \cite{Aghababaie:2003wz}.

The six-dimensional theory of the original Salam-Sezgin model suffers from anomalies. However generalised constructions can be made giving anomaly-free extensions of the Salam-Sezgin model \cite{RandjbarDaemi:1985wc, Ferrara:1996wv}. Many of these anomaly-free models can be related to Ho\v{r}ava-Witten type constructions  \cite{Horava:1996ma} producing a class of 7D/6D models with gauged R-symmetries, yielding in turn anomaly-free six-dimensional models on the boundary of the seven-dimensional space \cite{Pugh:2010ii}.

The original Salam-Sezgin model \cite{Salam:1984cj} consisted of six-dimensional $N=1$ supergravity coupled to one tensor and one vector multiplet and considered a dimensional reduction to four dimensions on a sphere and monopole background arising as a solution of the six-dimensional field equations. This work maintained that a consistent reduction of the six-dimensional theory on such a background can lead to a chiral theory in four dimensions. A detailed dimensional reduction on this background was later carried out in \cite{Gibbons:2003gp}, however, where it was found that, although the reduction does lead to Weyl fermions in the reduced theory, they are not gauge coupled in a complex representation of any gauge symmetry group and accordingly the reduced theory turns out to be non-chiral. The question of whether one may indeed obtain a chiral four-dimensional theory by dimensional reduction in this way thus remained open and will be the main focus of the present article.

The issue of chirality is often unclearly described in discussions about the intersection of particle physics and gravity, so it is worth being more specific about what constitutes a chiral theory, at least in the sense of particle physics. All Lagrangian field theories are PCT-invariant, so if a chiral spinor field carries a complex representation under some group $G$, its conjugate must carry the conjugate $G-$representation. An example of this is found in the rigid R-symmetry representations carried by spinors in many supersymmetric theories, \eg $N=4$ super Yang-Mills theory, where the left-handed spinors carry a fundamental {\bf 4} representation of $SU(4)$. This is not what would be considered a chiral theory in particle physics, however. One expression of this non-chiral structure is the fact that the $N=4$ theory can be re-expressed in terms of Majorana spinors without violating any of the essential physics of the theory -- in a Majorana formulation the R-symmetry reduces to $SO(4)$, but the essential physics is unchanged. The key indications of chiral structure lie in the interactions, not in free kinetic terms or in a summary account of fields and representations. Thus, another expression of the non-chiral nature of $N=4$ super Yang-Mills is the fact that the {\em gauge} couplings of the fermions are in the adjoint representation, which is real.

In characterising chiral theories such as the Standard Model, it is clearest to discuss chirality in terms of 4-component spinors. Then one often catalogues fields together with charge-conjugated fields of a single chirality, \eg left-handed $SU(2)$ lepton doublets together with the left-handed charge conjugates of right-handed $SU(2)$ lepton singlets. In a chiral theory analysed this way, the left-handed and the conjugated right-handed spinor species transform in inequivalent representations and have different couplings to other fields, \eg gauge and Yukawa couplings. Another chiral possibility is one where the left-handed  charge conjugates of right-handed spinors are simply absent, \eg for massless Standard Model neutrinos, which are then purely left-handed. In any case, the Standard Model is a chiral theory because the left-handed quarks and the left-handed charge-conjugates of right-handed quarks carry different gauge $SU(3)$ representations ($\bf3$ versus $\bar{\bf3}$), and so couple differently to the gluons -- as well as carrying different $SU(2)\times U(1)$ representations. The chiral nature of the Standard Model is also manifested in the Yukawa interactions, which involve the left-handed and left-handed conjugates of right-handed spinors quite asymmetrically.

If the left-handed and conjugated right-handed species have the same couplings, the theory is considered non-chiral or ``vectorlike''. Another way of characterising a non-chiral theory is one in which the spinors can all be rewritten as Majorana spinors without violating any gauge symmetries or other interaction structure.

For the above reasons, the non-gauged complex R-symmetry representations carried by chiral spinors in supersymmetric theories do not qualify such theories as chiral. It is also in this sense that the reduction of the Salam-Sezgin model from six to four dimensions considered in \cite{Gibbons:2003gp} did not generate a chiral theory. Although that sphere and monopole reduction remarkably generated a consistent $SU(2)$ gauging in four dimensions, the surviving fermions transformed either in the real {\bf 3} of $SU(2)$ or were singlets, so the four-dimensional theory turned out to be non-chiral.

The main aim of the present article is to lay out more fully the possibility of chiral reductions of theories such as the Salam-Sezgin model. In Section \ref{6D N=1 Supergravity}, we first outline six-dimensional $N=1$ supergravity, which will be our starting point for this discussion. Then in Section \ref{Reduction} we will demonstrate that a reduction on the Salam-Sezgin background to a chiral theory in four dimensions is in fact possible. This is because there is an interesting bifurcation in the reduction ansatz which means that two different consistent sets of four-dimensional excitations of the six-dimensional theory are possible. One of these consistent sets of excitations leads to the non-chiral theory with $SU(2)$ gauge group considered in \cite{Gibbons:2003gp} and the other, which we present here, leads to a chiral theory with a massive Stueckelberg gauged $U(1)_R$. The four-dimensional theory that we obtain in this way preserves the original chirality of the six-dimensional theory one began with. 

Both the six-dimensional theory that we consider in Section \ref{6D N=1 Supergravity} and the four-dimensional theory that we obtain in Section \ref{Reduction} contain Weyl fermions charged under a gauged $U(1)_R$ symmetry, making these theories chiral. However, in both cases this gauged $U(1)_R$ is the R-Symmetry of the theory -- either the gauged R-symmetry of the original model or of its dimensional reduction. This unusual situation renders the chirality found these cases rather irrelevant from a physical point of view. In order to obtain more physically relevant examples of chirality in the reduced theory, we consider the coupling of additional hypermultiplets in the starting six-dimensional theory. In Section \ref{HypermultipletCouplings}, we analyse these couplings and carry out the corresponding reduction on the Salam-Sezgin sphere and monopole background to give a four-dimensional theory with genuinely chiral fermions. Finally in Section \ref{GaugingHyperSymmetries}, we consider gauging the symmetries of the six-dimensional hypermultiplet by additional vector multiplets and analyse the resulting symmetries in the reduced four-dimensional theory.

\section{Six-dimensional $N=1$ Supergravity} 
\label{6D N=1 Supergravity}

We begin by considering the action for six-dimensional $N=1$ supergravity with field content $( \wh e_M{}^A, \wh \p_M, \wh B_{M N}^+ )$ coupled to one tensor multiplet $( \wh B_{MN}^- , \wh \c, \wh \f)$ and to one vector multiplet $(\wh A_M, \wh \l)$ gauging a $U(1)_R$ subgroup of the R-symmetry. Here $M = 0,\ldots , 5$ is a world index and $A= 0, \ldots , 5 $ is a tangent-space index. This action was also the starting point for the original Salam-Sezgin model \cite{Salam:1984cj}; it is given by
\ba
S &= \int d^6 x \wh e \bls \wh R - \fr14 \wh \pa_M \wh \f \wh \pa^M \wh \f - \fr1{12} e^{ \wh \f } \wh H_{MNR} \wh H^{MNR} - \fr14 e^{\fr12 \wh \f} \wh F_{MN} \wh F^{MN} - 8 g^2 e^{-\fr12 \wh \f} \nn \\
& \quad + \wh{\bar\p}_M \wh \G^{MNR} \wh D_N \wh \p_R + \wh{\bar \c} \wh \G^M \wh D_M \wh \c + \wh{\bar \l} \wh \G^M \wh D_M \wh \l   + \fr14 \big( \wh{ \bar \c} \wh \G^N \wh \G^M \wh \p_N + \wh {\bar \p}_N \wh \G^M \wh \G^N \wh \c  \big) \wh \pa_M \wh \f \nn \\
& \quad + \fr1{24} e^{\fr12 \wh \f } \wh H_{MNR} \big( \wh{\bar\p}{}^S \wh \G_{[S} \wh \G^{MNR} \wh \G_{T]} \wh \p{}^{T}  + \wh{\bar\p}_S \wh \G^{MNR} \wh \G^S \wh \c - \wh{\bar \c} \wh \G^S \wh \G^{M N R} \wh \p_S - \wh{\bar \c} \wh \G^{MNR} \wh \c + \wh{\bar\l} \wh \G^{MNR} \wh \l \big) \nn \\
& \quad - \fr{1}{4 \sq{2} } e^{\fr14 \wh \f} \wh F_{M N} \big( \wh{\bar \p}_R \wh \G^{M N} \wh \G^R \wh \l + \wh{ \bar \l} \wh \G^R \wh \G^{M N } \wh \p_R + \wh{\bar \c} \wh \G^{MN} \wh \l - \wh{ \bar \l } \wh \G^{MN} \wh \c \big) \nn \\
& \quad + i \sq{2} g e^{- \fr14 \wh \f} \big( \wh {\bar\p}_M \wh \G^M \wh \l + \wh{\bar\l} \wh \G^M \wh \p_M - \wh{\bar \c} \wh \l + \wh{\bar\l} \wh \c \big) \brs\ , 
\label{6DAct}
\ea 
where $\wh D_M \wh \c = \wh \pa_M \wh \c + \fr14 \wh \o_{MAB} \wh \G^{AB} \wh \c - i g \wh A_M  \wh \c$ and similarly for derivatives acting on $\wh \l$ and $\wh \p_M$. Here $M = 0, \ldots 5 $ is a world index raised and lowered with the metric $\wh g_{MN}$ and $A = 0, \ldots 5$ is a tangent-space index raised and lowered with $\wh \h_{AB} = \text{diag} ( - , +, ..., +)$. The field strengths appearing in this action are defined by $\wh H_{MNR} = 3 \wh \pa_{[M} \wh B_{N R]} + \fr32 \wh F_{[M N} \wh A_{R]}$ and $\wh F_{MN} = 2 \wh \pa_{[M} \wh A_{N]}$. We use the conventions $\wh R_{MN} = \wh R^R{}_{M R N} $, $\wh R_{MN}{}^{AB} = 2 \wh \pa_{[M} \wh \o_{N]}{}^{AB} + 2\wh \o_{[M}{}^{ A C} \wh \o_{N]}{}_{C}{}^B  $, $\wh {\bar \c} = i \wh \c^{\dagger} \wh \G_0$ .  The fermions considered here have chiralities such that $\wh \G_7 \wh \l  = \wh \l  $, $\wh \G_7 \wh \c = - \wh \c$, and $ \wh \G_7 \wh \p_A = \wh \p_A$ .  

This action is supersymmetric under the transformations
\ba
\d \wh e_M{}^A &= - \fr14 \wh {\bar \e} \wh \G^A \wh \p_M + \fr14 \wh {\bar \p}_M \wh \G^A \wh \e\ , &
\delta\wh{\psi}_M &=
\wh D_M\, \wh \e  + \fr1{48} e^{\fr12\wh \phi} \,
\wh H_{NPQ}\,\wh\Gamma^{NPQ}\, \wh\Gamma_M\, \wh\e\ ,\nn\\
\d \wh \f &= \fr12 \wh {\bar \e} \wh \c + \fr12 \wh {\bar \c} \wh \e\ , & 
\delta\wh\chi &=
-\fr14 ( \pa_M\wh\phi\, \wh\Gamma^M - \fr16 e^{\fr12\hat\phi} \, 
\wh H_{MNP}\, \wh\Gamma^{MNP}) \wh\e\ , \nn \\
\d \wh{A}_M &= \fr1{2 \sq{2} } e^{-\fr14 \wh \f} ( \wh {\bar \e} \wh \G_M \wh \l - \wh {\bar \l} \wh \G_M \wh \e)\ , &
\delta \wh\lambda &=
\fr1{4\sqrt{2}} ( e^{\fr14\wh\phi}\, \wh F_{MN}\, \wh\Gamma^{MN} - 8 i \,
g\, e^{-\fr14\wh \phi} ) \wh\e\ , \nn \\
\d \wh B_{M N}  &= \wh  A_{[M} \d \wh A_{N]}  + \fr14 e^{-\fr12 \wh \f} \big(  2 \wh {\bar \e} \wh \G_{[M} \wh \p_{N]}  + 2 \wh {\bar \p}_{[M} \wh \G_{N]} \wh \e + \wh {\bar \e} \wh \G_{M N } \wh \c - \wh {\bar \c} \wh \G_{MN} \wh \e \big) \ , \hspace{-6 cm} & 
\label{6DSusy}
\ea
where the supersymmetry parameter $\wh \e$ satisfies $\wh \G_7 \wh \e =  \wh \e$.

\section{Reduction on a Sphere and Monopole Background}
\label{Reduction}

\subsection{Reduction of the Bosonic Sector}
\label{Reduction of the Bosonic Sector}

We now consider a reduction of the above action on the background originally considered by Salam and Sezgin \cite{Salam:1984cj}. This reduction will be carried out at the level of the field equations which will then be integrated to give the reduced action. In many, but not all cases \cite{Lu:2006dh}, consistency of a dimensional reduction scheme may be checked either by direct verification in the field equations or by verification that insertion of the ansatz into the action commutes with variation of the action. However, the fundamental definition of a consistent Kaluza-Klein reduction is one in which solutions to the reduced lower-dimensional field equations yield exact solutions to the higher-dimensional equations when they are re-expressed via the reduction ansatz as higher-dimensional fields. 

We therefore begin by varying the six-dimensional action \eqref{6DAct} to find the field equations for the bosonic sector; neglecting bifermionic terms these read
\ba
\wh R_{M N} &= \fr14 \pa_M \wh \f \pa_N \wh \f + \fr12 e^{\fr12 \wh \f} ( \wh F_{M R} \wh F_N{}^R - \fr18 \wh F_{RS} \wh F^{RS} \wh g_{MN}) \nn \\
& \quad + \fr14 e^{ \wh \f } ( \wh H_{M RS} \wh H_{N}{}^{RS} - \fr16 \wh H_{RST} \wh H^{RST} \wh g_{MN} ) + 2 g^2 e^{-\fr12 \wh \f } \wh g_{MN}\ , \nn  \\
d \wh * d \wh \f &=  - \fr12 e^{\fr12 \wh \f}  \wh * \wh F_{\tw} \wedge \wh F_\tw  - e^{\wh \f} \wh * \wh H_\th \wedge \wh H_\th + 8g^2 e^{-\fr12 \wh \f} \wh * 1\ ,  \nn \\
d ( e^{\fr12 \f} \wh * \wh F_\tw ) &= e^{\wh \f} \wh * \wh H_\th \we \wh F_\tw\ , \hspace{2cm} d ( e^{\wh \f} \wh * \wh H_\th ) = 0\ , 
\ea
where the field strengths considered satisfy the Bianchi identities
\ba
d \wh H_\th &= \fr12 \wh F_\tw \we \wh F_\tw\ , & d \wh F_\tw &= 0\ . 
\label{bianchi1}
\ea

The Salam-Sezgin background \cite{Salam:1984cj} solving these equations is a product of four-dimensional Minkowski space and a 2-sphere with a monopole on it. An ansatz for a consistent set of four-dimensional fluctuations about this background was given in \cite{Gibbons:2003gp}. This leads, however, to a four-dimensional theory without chirality. This motivates us to try to find an alternative consistent ansatz leading to a chiral theory. With this in mind, and by a process of trial and error, we find the ansatz
\ba
\wh F_\tw &= \fr1{2g} \O_\tw - F_\tw\ , \nn \\
\wh H_\th &= H_\th - \fr{1}{2g} A_\on \we \O_\tw\ ,& 
\wh \f &= \vf - \f \ ,\nn \\
\wh e^\a &= e^{\fr14 ( \f + \vf) } e^\a\ , &
\wh e^a &= e^{ - \fr14 (\f + \vf) } e^a\ ,
\label{BosonicAnsatz}
\ea 
where $\O_2 = \sin y^5 d y^5 \we d y^6 = 4g^2 \e_{a b} e^a \we e^b$ is the volume form for the unit sphere $\e_{56}  = 1$ and where the sphere's vielbein $e^a$ has an associated curvature tensor such that $R_{m n} = 8 g^2 g_{m n}$ and $F_\tw = d A_\on$.  Here we have split the six-dimensional world index $M$ into $\m = 0, \ldots 3 $ and $m = 5, 6$ and similarly split the tangent space index $A$ into $ \a = 0 , \ldots 3$ and $a = 5, 6$. This is thus an ansatz for a reduction on the same background as that in \cite{Gibbons:2003gp} but with a different set of fields retained in the reduced theory. One major difference between these two sets of fluctuations is that there is an extra degree of freedom here, associated with the difference between the dilaton and the Kaluza-Klein scalar, whereas in \cite{Gibbons:2003gp} there were additional degrees of freedom associated with the Kaluza-Klein vector resulting in an $SU(2)$ Yang-Mills gauging in four dimensions. It is important to note that neither this ansatz nor the ansatz of Ref.\ \cite{Gibbons:2003gp} can be truncated into the other, implying that there is a bifurcation into two different branches of consistent excitations about the Salam-Sezgin background. 

Substituting the ansatz \eqref{BosonicAnsatz} into the Bianchi identities \eqref{bianchi1} implies
\ba
d H_\th &= \fr12 F_\tw \we F_\tw\ , & 
d F_\tw &= 0\ .
\ea
We next note that
\ba
\wh * \wh F_\tw &= 4 g e^{\fr32 (\f + \vf) } * 1 - \fr1{8g^2} e^{-\fr12 (\f + \vf) } * F_\tw \we \O_\tw\ , &
\wh * 1 &= \fr1{8g^2} e^{\fr12 (\f + \vf) } * 1 \we \O_\tw\ , \nn \\
\wh * \wh H_\th &= \fr{1}{8g^2} e^{- ( \f + \vf) } * H_\th \we \O_\tw - 4 g e^{ \f + \vf } * A_\on\ , &
\wh * d \wh \f &= \fr1{8 g^2} * d ( \vf - \f ) \we \O_\tw\ . 
\ea
We can then use these identities in the field equations of the six-dimensional bosonic fields in order to obtain reduced field equations describing the four-dimensional fluctuations. In this way, we find that the $\wh H_\th$ field equation leads to the two four-dimensional equations
\ba
d( e^{- 2\f} * H_\th) &= 0\ , &
d (e^{2\vf} * A_\on ) &= 0\ .
\label{4Dbianchi}
\ea
Similarly substituting the ansatz into the $\wh F_\tw$ field equation, we find
\ba
d ( e^{ \f} * F_\tw )  &= e^{- 2 \f} * H_\th \we F_\tw + 16 g^2 e^{2\vf}  * A_\on\ , 
\ea
while substituting  into the $\wh \f$ field equation gives 
\ba
d * d \f - d*d \vf &=  e^{- 2\f} * H_\th  \we H_\th+ \fr12 e^{-\f} * F_\tw \we F_\tw\ ,\nn \\
& \quad + 16 g^2 e^{2\vf} * A_\on \we A_\on - 8 g^2 e^\f (1 - e^{2 \vf} ) * 1\ .
\label{4DphiEOM1}
\ea
When considering the six-dimensional metric's field equations we must consider equations where the indices lie in the compact and non-compact directions independently. Considering the $\wh R_{ab}$ field equation, we find\footnote{
In deriving this we have used the lemma
\ba
\wh R_{\a\beta} &= e^{-\fr12 (\f + \vf) }\, \Big[ R_{\a\beta} - \fr14\,
\square(\f + \vf)\, \eta_{\a\beta} -\fr14 \pa_\a(\f + \vf)\, \pa_\beta(\f + \vf) 
\Big]\ ,  \nn\\
\wh R_{\a b} &= 0\ , \hspace{2cm} 
\wh R_{ab} = e^{\fr12\, (\f + \vf)}\, R_{ab} +\fr14 \, e^{-\fr12 (\f + \vf)}\, 
\square(\f + \vf) \, \delta_{ab} \ . \nn
\ea
}
\ba
d * d \f +  d*d \vf  &=  e^{- 2\f} * H_\th  \we H_\th+ \fr12 e^{-\f} * F_\tw \we F_\tw   \nn \\
& \quad - 16 g^2 e^{2\vf} * A_\on \we A_\on - 4 g^2 e^\f (2 - 8e^{ \vf}  + 6 e^{2 \vf} ) * 1\ ,
\label{4DphiEOM2}
\ea
where we have used $R_{m n} = 8 g^2 g_{m n}$.  Equations \eqref{4DphiEOM1} and \eqref{4DphiEOM2} can then be rearranged to give 
\ba
d * d \f  & =  e^{ - 2 \f} * H_\th  \we H_\th+ \fr12 e^{- \f} * F_\tw \we F_\tw  - 8 g^2 e^{ \f }  (1 - e^{ \vf} )^2 * 1\ , \nn \\
d * d \vf & = - 16 g^2 e^{2 \vf} * A_\on \we A_\on + 16 g^2 e^{\f} e^{ \vf} ( 1 - e^{\vf} ) * 1\ .
\ea
Substituting the ansatz into the $\wh R_{\a a}$ field equation gives an identity $0 = 0$, while the $\wh R_{\a \b}$ field equation and  \eqref{4DphiEOM2} together give 
\ba
R_{\a \b} &= \fr12 \pa_\a \f \pa_\b \f + \fr12 \pa_\a  \vf  \pa_\b \vf 
 + \fr12 e^{- \f} ( F_{\a \g} F_\b{}^\g - \fr14 F_{\g\d} F^{\g \d} \h_{\a \b} ) \nn \\
& \quad + \fr14 e^{- 2 \f} ( H_{\a \g \d} H_{\b} {}^{\g \d} - \fr13 H_{\g \d \l} H^{\g \d \l} \h_{\a \b} ) 
 + 8 g^2 e^{2 \vf}  A_\a A_\b  + 4 g^2 e^\f  ( 1 - e^{ \vf} )^2 \h_{\a \b}\ . 
\label{4DeEOM1}
\ea
Integrating these reduced field equations, we find the bosonic part of the four-dimensional reduced action
\ba
\cL_B &= R * 1 - \fr12 * d \f \we d \f - \fr12 * d \vf \we d \vf  - \fr12 e^{-2\f} * H_\th \we H_\th \nn \\
& \quad - \fr12 e^{-\f} *F_\tw \we F_\tw - 8 g^2 e^{2\vf} * A_\on \we A_\on - 8 g^2 e^\f ( 1 - e^{\vf} )^2 *1\ , 
\label{4DBosAct1}
\ea
where $F_\tw = d A_\on$, $H_\th = d B_\tw + \fr12 \o_\th$ and $d \o_\th = F_\tw \we F_\tw$ .

The 3-form field strength $H_\th$ appearing in this action can be dualised to the gradient of an axionic scalar. This is done in the standard way by adding a term to the action containing a Lagrange multiplier that imposes the Bianchi identity \eqref{4Dbianchi}
\be
\cL^\pr = - \s ( dH_\th - \fr12 F_\tw \we F_\tw) \ . 
\ee
Varying the action with respect to $H_\th$ then gives
\be
H_\th = e^{2\f} * d \s\ , 
\label{dualH}
\ee
and substituting this back into \eqref{4DBosAct1} gives
\ba
\cL_B &= R * 1 - \fr12 * d \f \we d \f - \fr12 * d \vf \we d \vf  - \fr12 e^{2\f} * d \s \we d \s - \fr12 e^{-\f} *F_\tw \we F_\tw \nn \\
&\quad  - 8 g^2 e^{2\vf} * A_\on \we A_\on + \fr12 \s F_\tw \we F_\tw - 8 g^2 e^\f ( 1 - e^{\vf} )^2 *1\ . 
\label{4DBosAct2}
\ea

\subsection{Reduction of the Supersymmetry Transformations}
\label{ReductionSusy}

In \cite{Gibbons:2003gp} it was noted that substituting the solution for the background we are considering into the six-dimensional supersymmetry transformations implies that the six-dimensional spinors should be expanded with respect to a background spinor $\h$ which is required to obey  $(\na_a - i g A_{a}^{\text{mono}} )\h = 0$, where $ d A^{\text{mono}}_{\on} = \fr1{2g} \O_\tw $ and $\s_3 \h = \h$. We chose a normalisation for this background spinor where $\bar \h \h = 1$ . Here we have decomposed the six-dimensional gamma matrices as
\ba
\wh \G_\a &= \g_\a \ct \s_3\ , &  \wh \G_a &= \id \ct \s_a\ , 
\ea
where $\s_a$, $\s_3$ are Pauli matrices.  This implies, 
\be
\wh \G_7 = \g_5 \ct \s_3\ . 
\ee
Using the background spinor, we make the following ansatz, again obtained by a process of trial and error, for the fluctuations in the six-dimensional fermions
\ba
\wh \p_\a &= e^{ - \fr18 ( \f + \vf)} \left[ \p_\a \ct \h + \fr1{2\sq 2} \g_\a ( \c + \z) \ct \h \right]\ , &
\wh \l &= e^{ - \fr18 ( \f + \vf)} \l \ct \h\ ,\nn \\
\wh \p_a &= - \fr1{ 2 \sq 2} e^{ - \fr18 ( \f + \vf)} ( \c + \z ) \ct \s_a \h\ , &
\wh \c &= \fr1{\sq 2} e^{ - \fr18 ( \f + \vf)} ( \c - \z ) \ct \h\ ,\nn \\
\wh \e &=  e^{ \fr18 ( \f + \vf)} \e \ct \h\ . 
\label{FermionicAnsatz}
\ea
The chirality properties of the six-dimensional fermions then imply the inherited four-dimensional chiralities $ \g_5 \l = \l$,  $\g_5 \c = - \c$,  $\g_5 \z = - \z$, $\g_5 \p_\a = \p_\a$, and $\g_5 \e = \e$. Substituting the ansatz \eqref{FermionicAnsatz} and \eqref{BosonicAnsatz} into the supersymmetry transformations of the six-dimensional spinors \eqref{6DSusy}, we obtain the supersymmetry transformations for the four-dimensional spinors appearing in the ansatz\footnote{
We have made use of the lemmas
\ba
\wh \omega_{\a\beta} &=\omega_{\a\beta} + \fr14 \,e^{-\fr14(\f +\vf)} \,  
(\pa_\b (\f + \vf) \, \hat e_\a - \pa_\a (\f + \vf) \, \hat e_\beta)\ ,\nn\\
\hat\omega_{\a b} &= \fr14 \, e^{-\fr14 ( \f + \vf) }\, \pa_\a ( \f + \vf) \, \hat e_b\ , \hspace{2cm}
\hat\omega_{ab} = \omega_{ab}\ . \nn
\ea
}
\ba
\d \p_\a &= \na_\a \e + i g ( 1- e^\vf ) A_\a \e - \fr1{24} e^{-\f} H_{\b\g\d} \g_\a{}^{\b\g\d}\e\ ,  \nn \\
\d \l &= - \fr1{4 \sq{2} } e^{ - \fr12 \f} F_{\a \b} \g^{\a \b} \e + \sq{2} i g e^{\fr12 \f} ( e^{ \vf } - 1) \e\ ,\nn \\
\d \z &= \fr1{2\sq{2}} \pa_\a \vf \g^\a \e + i g \sq 2 e^\vf A_\a \g^\a \e\ ,\nn \\
\d \c &= \fr1{2\sq{2}} \pa_\a \f \g^\a \e + \fr1{12\sq 2} e^{-\f} H_{\a\b\g} \g^{\a\b\g} \e\ . 
\ea
In a discussion of chirality in the reduced theory, the crucial factor is whether any of the Weyl fermions appearing in this reduction are charged under a complex representation of some group. The gauged $U(1)_R$ present in the bosonic part of the reduced theory \eqref{4DBosAct2} is broken by a mass term. However it is possible to form a Stueckelberg extension of the ansatz leading to a reduced theory in which the gauged $U(1)_R$ is unbroken. To do this, we modify the the ansatz for $\wh H_{\a a b}$ to read
\ba
\wh H_{\a ab} &=  - 4 g e^{\fr14 (\f + \vf)} ( \pa_\a \r + A_\a ) \e_{a b}\ ,
\label{StueckelbergAnsatz}
\ea
where we have introduced a Stueckelberg scalar $\r$, while keeping the ansatz for the vector unchanged, $\wh A_\a = - e^{- \fr14 (\f - \vf) } A_\a$. Carrying this out and dualising the 3-form using \eqref{dualH}, we obtain the supersymmetry transformations of fermions in the reduced theory:
\ba
\d \p_\a &= \na_\a \e + i g ( 1- e^\vf ) A_\a \e - \fr1{4} i e^{\f} \pa_\a \s \e\ , \nn \\
\d \l &= - \fr1{4 \sq{2} } e^{- \fr12 \f} F_{\a \b} \g^{\a \b} \e + \sq{2} i g e^{\fr12 \f} ( e^{ \vf } - 1) \e\ ,\nn  \\
\d \z &= \fr1{2\sq{2}} \pa_\a \vf \g^\a \e + i g \sq{2} e^\vf D_\a \r \g^\a \e\ ,  \nn \\
\d \c &= \fr1{2\sq{2}} \pa_\a \f \g^\a \e + \fr1{2\sq{2}} i e^{\f} \pa_\a \s \g^\a \e\ .
\ea
We now note that the fields of the reduced theory form the supergravity multiplet $(e_\m{}^\a, \p_\m)$, a scalar multiplet $(\f, \s, \c)$ and a massive Stueckelberg vector multiplet $(\vf, \r, \z, A_\m, \l )$.  

Next, we examine the reduction of the supersymmetry transformations of the bosonic fields. Considering the transformation of $\wh e_m{}^a$, we find the need to carry out a compensating Lorentz transformation which acts as $\d \wh e_M{}^A = \L^A{}_B \wh e_M{}^B$, where 
\be
\L^a{}_b = - \fr1{8\sq2} i \e^a{}_b ( \bar \e \c - \bar \c \e + \bar \e \z - \bar \z \e )\ , 
\ee
in order to preserve the Lorentz gauge choice made in the ansatz for $\wh e^A$.
Then the modified transformation of $\wh e_m{}^a$ combined with the transformation of $\wh \f$ implies
\ba
\d \f &= - \fr1{2 \sq 2}  ( \bar \e \c + \bar \c \e)\ ,  &
\d \vf &= - \fr1{2 \sq 2}  ( \bar \e \z + \bar \z \e)\ . 
\ea
Similarly, the transformation of $\wh A_\m$ implies
\be
\d A_\m = - \fr1{2 \sq{2} } e^{\fr12 \f} ( \bar \e \g_\m \l - \bar \l \g_\m \e)\ . 
\ee
Finally, making an additional compensating Lorentz transformation with parameter 
\be
\L^\a{}_\b = \fr1{8\sq2} ( \bar \e \g^{\a}{}_\b \c - \bar \c \g^\a{}_\b \e +\bar \e \g^{\a}{}_\b \z - \bar \z \g^\a{}_\b \e )\ ,
\ee 
in order to bring the four-dimensional supersymmetry transformations into the standard form, we find that the variation of the vielbein in the reduced theory is given in the end by 
\be
\d e_\m{}^\a = - \fr14 ( \bar \e \g^{\a} \p_\m - {\bar \p}_\m  \g^\a \e )\ .
\ee

\subsection{Reduction of the Fermionic Sector} 
\label{ReductionFermionic} 
We now consider the reduction of the six-dimensional fermionic field equations
\ba
\wh \G^{MNR} \wh D_N \wh \p_R &= - \fr14 \wh \G^N \wh \G^M \wh \c \wh \pa_N \wh \f 
- \fr1{24} e^{\fr12 \wh \f} \wh H^{NRS} \big( \wh \G^{[M} \wh \G_{NRS} \wh \G^{T]} \wh \p_T   + \wh \G_{NRS} \wh \G^M \wh \c \big) \nn \\
&\quad 
+ \fr1{4 \sq{2} } e^{\fr14 \wh \f} \wh F_{NR} \wh \G^{NR} \wh \G^M \wh \l - i \sq{2} g e^{- \fr14 \wh \f} \wh \G^M \wh \l\ ,  \nn \\
\wh \G^M \wh D_M \wh \l &= - \fr1{24} e^{\fr12 \wh \f} \wh H_{MNR} \wh \G^{MNR} \wh \l 
+ \fr1{4 \sq{2} } e^{\fr14 \wh \f} \wh F_{MN} \big( \wh \G^R \wh \G^{MN} \wh \p_R 
- \wh \G^{MN} \wh \c \big)  \nn \\
&\quad
-  i \sq{2} g e^{- \fr14 \wh \f} \big( \wh \G^M \wh \p_M + \wh \c \big)\ , \nn \\
\wh \G^M \wh D_M \wh \c &= - \fr14 \wh \G^N \wh \G^M \wh \p_N \wh \pa_M \wh \f 
+ \fr1{24} e^{\fr12 \wh \f} \wh H_{MNR} \big(  \wh \G^S \wh \G^{MNR} \wh\p_S + \wh \G^{MNR} \wh \c \big) \nn \\
&\quad
+ \fr1{4 \sq{2} } e^{\fr14 \wh \f} \wh F_{MN} \wh \G^{MN} \wh \l 
+ i \sq{2} g e^{- \fr14 \wh \f} \wh \l\ .
\ea 
As before, we obtain the field equations describing the consistent sets of fluctuations about the background by substituting our ansatz \eqref{BosonicAnsatz} and \eqref{FermionicAnsatz} into the six-dimensional field equations. Carrying this out for the the $\wh \l $ field equation gives
\ba
 \g^\a &( \na_\a + i g A_\a ) \l =\nn \\
& 
- \fr1{24} e^{- \f} H_{\a \b \g} \g^{\a\b\g}  \l + i g e^\vf A_\a \g^\a \l 
- \fr1{4 \sq{2} } e^{-\fr12 \f} F_{\a\b} \big( \g^\g \g^{\a \b} \p_{\g} - \sq 2 \g^{\a\b} \c \big) \nn \\&
+ i \sq 2 g e^{\fr12 \f} ( e^\vf - 1) \big( \g^\a\p_\a + \sq 2 \c ) + i 4  g e^{\fr12 \f} e^\vf  \z\ .
\label{RedFermEOM1}
\ea
Similarly the $\wh \c$ field equation gives
\ba
\fr1{\sq 2} \g^\a &( \na_\a + i g A_\a) ( \c - \z ) =\nn \\
&
- \fr14 \g^\a \g^\b \p_\a \pa_\b ( \vf - \f ) 
+ \fr1{12 \sq 2} e^{-\f} H_{\a\b\g} \big( \sq 2 \g^\d \g^{\a\b\g} \p_\d + \g^{\a\b\g} \z + 3 \g^{\a\b\g} \c \big) \nn \\&
- i \fr1{\sq2} g e^\vf A_\a \big( \sq 2 \g^\b \g^\a \p_\b -\g^\a \c - 3 \g^\a \z \big) - \fr1{ 4 \sq 2} e^{- \fr12 \f} F_{\a\b} \g^{\a\b} \l \nn \\&
+ i \sq 2 g e^{\fr12 \f} ( e^\vf + 1) \l\ .
\ea
Substituting the ansatz into the $\wh \p_M$ field equation with the free index in the $\alpha$ direction gives
\ba
 \g^{\a \m \n} &( \na_\m + i  g A_\m  ) \p_\n 
 - \fr1{4\sq 2}  \g^\b \g^\a ( \z + \c ) \pa_\b ( \f + \vf) =\nn \\
 &  - \fr1{4\sq	2}  \g^\b \g^\a ( \c - \z ) \pa_\b ( \vf - \f)
 + \fr1{4} e^{- \f} H^{\a \b \g} \g_\g \p_\b - \fr1{12 \sq 2 } e^{-\f} H_{\b\g\d} \g^{\b\g\d} \g^\a \c \nn \\ &
  - i g  e^\vf A_\g \big( \g^{\a\b\g} \p_\b + \sq 2 \g^\g \g^\a \z \big)
 - \fr1{4 \sq 2} e^{- \fr12 \f } F_{\b\g} \g^{\b\g} \g^\a \l  + i \sq 2 g e^{\fr12 \f } ( e^\vf - 1 ) \g^\a \l\ .
\ea
The $\wh \p_M$ field equation with the free index in the $a$ direction gives
\ba
 \g^{\m \n} &( \na_\m + i g A_\m) \p_\n +  \fr1{\sq2} \g^\m ( \na_\m + i g A_\m) ( \c + \z)
- \fr14 \pa_\n ( \f + \vf)  \g^\m \g^\n \p_\m + \fr1{4\sq 2} \pa_\m ( \f + \vf) \g^\m ( \z + \c)= \nn \\
&  \fr1{4\sq2} \g^\a \pa_\a ( \vf - \f ) ( \c - \z ) 
- \fr1{12 \sq 2} e^{-\f} H_{\a\b\g} \big(\sq 2 \g^{\a \b\g\d} \p_\d - \g^{\a\b\g} \c + \g^{\a\b\g} \z \big) \nn \\ &
+ \fr{i}{\sq 2}  g e^\vf A_\a \big(\sq 2 \p^\a  + \g^\a \c - \g^\a \z \big) 
- \fr1{4 \sq 2} e^{- \fr12 \f } F_{\a \b} \g^{\a\b} \l - i \sq 2 g e^{\fr12 \f} ( e^\vf +1) \l\ . 
\label{RedFermEOM4} 
\ea
Equations \eqref{RedFermEOM1} to \eqref{RedFermEOM4} can then be rearranged into the field equations of the four-dimensional fermions given by
\ba
\g^{\m \n \r} D_\n \p_\r &= 
 \fr1{2\sq 2} \g^\n \g^\m \c \pa_\n  \f +  \fr1{2\sq 2} \g^\n \g^\m \z \pa_\n  \vf \ 
 + \fr1{4} e^{- \f} H^{\m \n \r} \g_\r \p_\n - \fr1{12 \sq 2} e^{-\f} H_{\r \s \t} \g^{\r\s\t} \g^\m \c \nn \\ 
 & \quad
  - i g e^\vf A_\r \big( \g^{\m\n\r} \p_\n + \sq 2 \g^\r \g^\m \z \big) 
 - \fr1{4 \sq 2} e^{- \fr12 \f } F_{\n \r} \g^{\n \r} \g^\m \l  + i \sq 2 g e^{\fr12 \f } ( e^\vf - 1 ) \g^\m \l\ , \nn \\
\g^\m D_\m \l &= 
- \fr1{24} e^{- \f} H_{\m \n \r} \g^{\m\n\r}  \l + i g e^\vf A_\m \g^\m \l 
- \fr1{4 \sq{2} } e^{-\fr12 \f} F_{\m\n} \big( \g^\r \g^{\m\n} \p_{\r} - \sq 2 \g^{\m\n} \c \big) \nn \\
&\quad
+ i \sq 2 g e^{\fr12 \f} ( e^\vf - 1) \big( \g^\m\p_\m + \sq 2 \c ) + i 4  g e^{\fr12 \f} e^\vf  \z\ , \nn\\
\g^\m D_\m \z &= 
\fr1{2 \sq 2} \g^\m \g^\n \p_\m \pa_\n \vf
- \fr1{24} e^{-\f} H_{\m\n\r} \g^{\m \n\r} \z
+ i  g e^\vf A_\m (  \sq 2 \g^\n \g^\m \p_\n  - 3 \g^\m \z ) 
- i 4 g e^{\fr12 \f } e^\vf \l\ , \nn \\
\g^\m D_\m \c &= \fr1{2 \sq 2}  \g^\m \g^\n \p_\m \pa_\n \f 
+ \fr1{24} e^{-\f} H_{\m\n\r} \big( \sq 2 \g^\s \g^{\m\n\r} \p_\s + 3 \g^{\m \n\r} \c \big) 
+ i g e^\vf A_\m \g^\m \c \nn \\
& \quad
- \fr1{4 } e^{-\fr12 \f} F_{\m \n} \g^{\m \n} \l 
- i  2 e^{\fr12 \f } ( e^\vf - 1 ) \l\ , 
\ea
where $D_\m \c = \pa_\m \c + \fr14 \o_{\m \a\b} \g^{\a\b}\c + i g A_\m \c$ and similarly for $\p_\m$, $\l$ and $\z$. 

Note that upon moving to the Stueckelberg version of the ansatz \eqref{StueckelbergAnsatz},  $ A_\m$ on the RHS of these equations becomes modified to $ \pa_{\m} \r +  A_\m$. However the $A_\m$ appearing in the covariant derivative $D_\m$ on the LHS of these equations is unmodified, so the derivative remains $U(1)_R$ covariantised. All spinors then carry the same charge under the local $U(1)_R$. 

Dualising  $H_{\m\n\r}$ \eqref{dualH}, moving to the Stueckelberg version of the ansatz \eqref{StueckelbergAnsatz}, integrating these field equations and combining with the bosonic part of the action derived in Section \ref{Reduction of the Bosonic Sector}, we find the full four-dimensional action
\ba
S &= \int d^4 x e \bls R - \fr12 \pa_\m \f \pa^\m \f - \fr12 \pa_\m \vf \pa^\m \vf - \fr12 e^{2\f} \pa_\m \s \pa^\m \s \nn \\
&\quad 
- 8 g^2 e^{2\vf} D_\m \r D^\m \r   - \fr14 e^{-\f} F_{\m \n} F^{\m \n}  - \fr18 \s \e^{\m\n\r\s} F_{\m \n} F_{\r \s}  - 8 g^2 e^\f ( 1-e^\vf)^2 \nn \\
&\quad
+ \bar \p_\m \g^{\m \n \r} D_{\n} \p_{\r} + \bar \l \g^\m D_\m \l +  \bar \c \g^\m D_\m \c +  \bar \z \g^\m D_\m \z \nn \\
&\quad
 - \fr1{2 \sq{2} } \big( \bar \p_\m \g^\n \g^\m \c + \bar \c \g^\m \g^\n \p_\m \big) \pa_\n \f 
- \fr1{2 \sq{2} } \big( \bar \p_\m \g^\n \g^\m \z + \bar \z \g^\m \g^\n \p_\m \big) \pa_\n \vf \nn \\ 
&\quad
+ \fr1{4} i e^{\f} \pa_{\m} \s \big ( \bar \p_\n \g^{\m \n \r} \p_\r + \sq 2 \bar \p_\n \g^{\m} \g^\n \c - \sq 2 \bar \c \g^\n \g^{\m} \p_\n  + 3 \bar \c \g^{\m} \c  -  \bar \z \g^{\m} \z+  \bar \l \g^{\m} \l \big) \nn \\
&\quad
+ i g e^\vf D_\m \r \big ( \bar \p_\n \g^{\m \n \r} \p_\r + \sq 2 \bar \p_\n \g^\m \g^\n \z - \sq 2 \bar \z \g^\n \g^\m \p_\n  + 3 \bar \z \g^\m \z  -  \bar \c \g^\m \c - \bar \l \g^\m \l \big) \nn \\
&\quad
+ \fr1{8 } e^{-\fr12 \f } F_{\m \n} \big( \sq 2 \bar \p_\r \g^{\m \n } \g^\r \l + \sq 2 \bar \l \g^\r \g^{\m \n} \p_\r + 2 \bar \c \g^{\m \n} \l - 2 \bar \l \g^{\m \n } \c  \big) \nn \\
&\quad 
- i g e^{\fr12 \f } ( e^\vf - 1) \big(\sq 2 \bar \p_\m \g^\m \l + \sq 2 \bar \l \g^\m \p_\m +  2 \bar \l \c - 2 \bar \c \l\big) 
- i  4  g e^{\fr12 \f } e^\vf ( \bar \l \z - \bar \z \l ) \brs\ , 
\ea
where $D_\m \r = \pa_\m \r + A_\m$. This is supersymmetric under 
\ba
\d e_\m{}^\a &= - \fr14 ( \bar \e \g^{\a} \p_\m - {\bar \p}_\m  \g^\a \e )\ , &
 \d \p_\a &= D_\a \e - i g e^\vf D_\a \r \e - \fr1{4} i e^{\f} \pa_\a \s \e\ ,\nn \\
\d A_\m &= - \fr1{2 \sq{2} } e^{\fr12 \f} ( \bar \e \g_\m \l - \bar \l \g_\m \e)\ ,\nn &
\d \l &= - \fr1{4 \sq{2} } e^{- \fr12 \f} F_{\a \b} \g^{\a \b} \e + \sq{2} i g e^{\fr12 \f} ( e^{ \vf } - 1) \e \nn \\ 
\d \vf &= - \fr1{2\sq{2}} ( \bar \e \z + \bar \z \e)\ ,&
\d \z &= \fr1{2\sq{2}} \pa_\a \vf \g^\a \e + i g \sq{2} e^\vf D_\a \r \g^\a \e\ ,\nn \\
\d \f &= - \fr1{2\sq{2}} ( \bar \e \c + \bar \c \e)\ ,  & 
\d \c &= \fr1{2\sq{2}} \pa_\a \f \g^\a \e + \fr1{2\sq{2}} i e^{\f} \pa_\a \s \g^\a \e\ , \nn \\
\d \s &=  \fr1{2\sq{2}} i e^{ - \f} ( \bar \e \c - \bar \c \e)\ ,& 
\d \r &=   \fr1{8\sq{2}g} i e^{ - \vf} ( \bar \e \z - \bar \z \e )\ ,
\ea
where the transformations of $\s$ and $\r$ are obtained by demanding closure of the supersymmetry algebra.

The reduced theory shown here is chiral because the Weyl fermions appearing in the action are charged under the $U(1)_R$ symmetry.  This shows that reduction of the Salam-Sezgin background can indeed preserve the chirality of the six-dimensional theory. However, as the chirality in this four-dimensional theory is caused only by couplings to a Stueckelberg-compensated gauged massive $U(1)_R$ R-symmetry, it is not very persuasive as a physical example of chirality preservation. In order to arrive at a four-dimensional theory with a more physical realisation of chirality we next will restart our discussion with hypermultiplets coupled to the six-dimensional theory. 

\section{Hypermultiplet Couplings}
\label{HypermultipletCouplings}
The fermions of the six-dimensional action \eqref{6DAct} that we have considered so far are six-dimensional Weyl fermions. However, another option in six dimensions is to write the action in terms of symplectic Majorana-Weyl fermions. This has not been done so far as it makes manifest an $Sp(1) \cong SU(2)$ symmetry which is broken in the reduction to four dimensions, and so adds an unnecessary additional level of complexity to the discussion. However, the couplings of hypermultiplets which we will now consider are intimately related to this $Sp(1)$ symmetry \cite{Nishino:1986dc, Nishino:1997ff} and it is therefore helpful to rewrite the bulk action that we considered before in terms of symplectic Majorana-Weyl spinors. To do this, we define 
\ba
\wh \c^{1} &=  \wh \c\ , &\wh \c^{ 2} &= i \wh C^{-1} \wh \G_0^T \wh{\c}{}^*\ . 
\label{6DSymMaj}
\ea
which implies 
\ba 
\wh{ \bar \c}{}^{\un A} \equiv i  ( \wh \c_{\un A} )^\dagger \wh \G_0 = ( \wh{ \c}{}^{\un A} )^T \wh C\ , 
\ea 
where $\wh C$ is the charge conjugation matrix with $\wh C \wh \G_A \wh C^{-1} = - \wh \G_A^T$,  $\wh C = \wh C^T$, $\un A = 1,2 $ is an $Sp(1)$ doublet index raised and lowered with the antisymmetric tensor $\e^{\un{A}\un{B}}= - \e^{\un{B} \un{A}}$, $\e_{12} = \e^{12} = 1$, such that $\c^\un A = \e^{\un A \un B} \c_\un B$, $ \c_\un B  = \c^\un A \e_{\un A \un B}$ and similarly for all other fermions in the 6D theory considered so far. 

The hypermultiplet fermions that we consider are not charged under the $Sp(1)$ but transform instead under an $Sp(n)$ with respect to which they are symplectic Majorana
\ba 
\wh{ \bar \p}{}^{\hat{\un a}} \equiv i  ( \wh \p_{\hat{\un a}} )^\dagger \wh \G_0 = ( \wh{ \p}{}^{\hat{\un a}} )^T \wh C\ , 
\ea 
where $\hat{\un a} = 1,..,2n$ is an $Sp(n)$ fundamental index which is raised and lowered with the $Sp(n)$ invariant tensor $\wh \e_{\hat{\un{a}} \hat{\un b} } = - \wh \e_{\hat{\un{b}} \hat{\un a}}$, $\wh \e_{\hat {\un{a}} \hat {\un c} } \wh \e^{\hat {\un{c}} \hat{\un b} } = - \wh \d_{\hat{\un a}}{}^{\hat{ \un b} }$,  $\p^\uh a = \e^{\uh a \uh b } \p_\uh b $, $ \p_\uh b  = \p^\uh a \e_{\uh a \uh b}$ . Here we will use the convention
\ba
\wh \e_{\uh a \uh b} = \left( \begin{array}{cc}
0 & \id_n \\
- \id_n & 0 
\end{array} \right) 
\ea 
where $\id_n$ is the $(n \times n)$ identity matrix.

The hypermultiplet scalars $\wh L_\uh \a{}^\uh a$  and $\wh L_\uh \a{}^{\un A} $ transform under a global left-acting $Sp(n,1)$ and a local right-acting $Sp(n) \otimes Sp(1)$. This means that their physical degrees of freedom describe the coset 
\ba
\fr{Sp(n,1)}{Sp(n)\ct Sp(1) }\ .
\ea
Here $\uh \a$ is an $Sp(n,1)$ index which is raised and lowered with the $Sp(n,1)$ invariant tensor $ \wh \O_{\hat{\un{\a}} \hat{\un \b} } = - \wh \O_{\hat{\un{\b}} \hat{\un \a}}$, $ \wh \O_{\hat {\un{\a}} \hat {\un \g} } \wh \O^{\hat {\un{\g}} \hat{\un \b} } = - \wh \d_{\hat{\un \a}}{}^{\hat{ \un \b} }$ and we use the convention 
\ba
\wh \O_{\uh \a \uh \b} = \left( \begin{array}{cccc}
0 & 1 & 0 & 0 \\
-1 & 0 & 0 & 0 \\
0 & 0 & 0 & \id_n \\
0 & 0 & - \id_n & 0 
\end{array} \right) .
\ea 
These scalars satisfy\footnote{The coupling to hypermultiplets to six-dimensional  $N=1$ supergravity was considered in \cite{Nishino:1986dc}. However the description of these couplings in terms of the coset representatives $\wh L^\uh \a{}_{\un A}$ and $\wh L^\uh \a{}_{\uh a}$ which has been carried out here is, to our knowledge, a new result. This alternate formulation is useful as it allows us to leave general the $U(1)_R$ gauging described by the generator $T_{\uh \a}{}^{\uh \b}$, which will later be constrained. }
\ba
\wh L^\uh \a{}_{\un A} \wh L_{\uh \a}{}^{\un B} &=  - \wh \d_\un A{}^\un B,  &
 \wh L^\uh \a{}_{\uh a}  \wh L_{\uh \a}{}^{\uh b} &=   \wh \d_\uh a{}^\uh b\ , \nn \\
\wh L^\uh \a{}_{\un A} \wh L_{\uh \a}{}^{\uh a} &= 0 \ , 
&  - \wh L_{\uh \a}{}^{\un A} \wh L^\uh \b{}_{\un A} + \wh L_{\uh \a}{}^{\uh a}  \wh L^\uh \b{}_{\uh a} &=   \wh \d_\uh \a{}^\uh  \b\ . 
\label{LIds}
\ea
and 
\ba
( \wh L_\uh \a{}^\uh a )^* &=  - \wh L^\uh \b{}_\uh a \wh \h_{\uh \b}{}^{\uh \a} &  ( \wh L_\uh \a{}^\un A )^* &=  - \wh L^\uh \b{}_\un A \wh \h_{\uh \b}{}^{\uh \a}
\ea
where $\wh \h_\uh \a{}^\uh \b$ is the $U(2n,2)$ invariant tensor
\ba
\wh \h_{\uh \a}{}^{\uh \b} &=   \left( \begin{array}{cc}
\id_2 & 0 \\
0 & -  \id_{2n} 
\end{array} \right)\ .
\ea
From these scalars, we build the Maurer-Cartan forms 
\ba
\wh P_{ M \uh a}{}^\un A &=  \wh L^\uh \a{}_{\uh a} \wh D_M \wh L_{\uh \a}{}^{\un A}\ , &
\wh Q_{M \un A}{}^\un B &= \wh L^\uh \a{}_{\un A }  \wh D_M \wh L_{\uh \a}{}^{\un B}\ ,  &
\wh Q_{ M \uh a}{}^\uh b &=  - \wh L^\uh \a{}_{\uh a} \wh D_M \wh L_{\uh \a}{}^{\uh b}\ , 
\ea
where\footnote{Note that here the generators are defined such that there is no factor of $i$ in the gauge covariantisation terms.}
\ba
\wh D_M \wh L_\uh \a{}^\un A &=\wh  \pa_M \wh L_\uh \a{}^{\un A} + g \wh A_M T_{\uh \a}{}^{\uh \b} \wh L_{\uh \b}{}^{\un A}\ ,   & 
\wh D_M \wh L_\uh \a{}^\uh a &= \wh  \pa_M \wh L_\uh \a{}^{\uh a} + g \wh A_M T_{\uh \a}{}^{\uh \b} \wh L_{\uh \b}{}^{\uh a}\ . 
\ea
When these hypermultiplets are included, the six-dimensional action \eqref{6DAct} becomes\ modified to 
\ba
S &= \int d^6 x \wh e \bls \wh R - \fr14 \wh \pa_M \wh \f \wh \pa^M \wh \f - 2 \wh P_{M \uh a \un A}  \wh P^{M \uh a \un A}  -   \fr1{12} e^{ \wh \f } \wh H_{MNR} \wh H^{MNR}  \nn \\
&\quad
- \fr14 e^{\fr12 \wh \f} \wh F_{MN} \wh F^{MN}  -   4 g^2 \wh C^{ \un {A} \un {B} }  \wh C_{\un{A} \un{B} } e^{-\fr12 \wh \f} \nn \\
&\quad
 + \fr12 \wh{\bar\p}^{\un A}_M \wh \G^{MNR} \wh D_N \wh \p_{R {\un A}} + \fr12 \wh{\bar \c}^{{\un A}} \wh \G^M \wh D_M \wh \c_{{\un A}}  + \fr12 \wh{\bar \l}^{{\un A}} \wh \G^M \wh D_M \wh \l_{{\un A}}  + \fr12 \wh{\bar \p}{}^{\hat{\un a}}  \wh \G^M \wh D_M \p_{\hat{\un a}}   \nn \\ 
&\quad
+ \fr14 \wh{ \bar \c}^{\un A} \wh \G^N \wh \G^M \wh \p_{N{\un A}}  \wh \pa_{M} \wh \f  -   \wh{ \bar \p}^{\hat{\un a}} \wh \G^N \wh \G^M \wh \p_{N}^{{\un A}}   \wh P_{ M \uh a \un A} \nn \\
&\quad
 + \fr1{48} e^{\fr12 \wh \f } \wh H_{MNR} \big( \wh{\bar\p}{}^{S{\un A}} \wh \G_{[S} \wh \G^{MNR} \wh \G_{T]} \wh \p{}^{T}_{\un A}  + 2 \wh{\bar\p}_S^{\un A} \wh \G^{MNR} \wh \G^S \wh \c_{\un A} - \wh{\bar \c}^{\un A} \wh \G^{MNR} \wh \c_{\un A} \nn \\
 &\quad
 + \wh{\bar\l}^{\un A} \wh \G^{MNR} \wh \l_{\un A}  -  \wh{ \bar \p}{}^{\hat{\un a}} \wh \G^{ MNR } \wh \p_{\hat{\un a}}\big) \nn \\
&\quad
 - \fr{1}{4 \sq{2} } e^{\fr14 \wh \f} \wh F_{M N} \big( \wh{\bar \p}_R^{\un A} \wh \G^{M N} \wh \G^R \wh \l_{\un A} + \wh{\bar \c}^{\un A} \wh \G^{MN} \wh \l_{\un A} \big) \nn \\
&\quad
-  2 \sq 2 g e^{ - \fr14 \wh \f}  \wh {\bar \p}{}^{\hat{\un a}} \wh \l^{ \un A} \wh \x_{\uh a \un A}  - \sq{2}  g e^{- \fr14 \wh \f} \wh C_{\un{A} \un{B}} \big(  \wh{\bar\l}{}^{\un A} \wh \G^M \wh \p_M^{\un B} + \wh{\bar\l}{}^{\un A} \wh \c^{\un B} \big)  \brs\ , 
\label{6DHyperAct}
\ea 
where
\ba
\wh D_M \wh \c^{ \un A} &= \wh \na_M \wh \c^{\un A}  +  \wh Q_{M}{}^{ \un A}{}_{\un B} \wh \c^{\un B}\ , & 
\wh C_{\un A}{}^{\un B} &= \wh L^{\uh  \a}{}_\un A T_{\uh \a}{}^{\uh \b} \wh L_{\uh \b}{}^{\un B} \ , \nn \\
\wh D_M \wh \p^{ \hat{\un a}} &= \wh \na_M \wh \p^{\hat{\un a}}  +  \wh Q_{M}{}^{ \uh a}{}_{\uh b} \wh \p^{ \hat{\un b }} \ , & 
\wh \x_{\uh a}{}^{\un A} &= \wh L^{\uh \a}{}_\uh a T_{\uh \a}{}^{\uh \b} \wh L_{\uh \b}{}^{\un A}\ . 
\ea
This action is supersymmetric under the transformations
\ba 
\d \wh e_M{}^A &= - \fr14 \wh {\bar \e}^{\un A}  \wh \G^A \wh \p_{M \un A}\ ,  &
\delta \wh{\psi}_M^{\un A}  &=
\wh D_M\, \wh \e^{\un A}   + \fr1{48} e^{\fr12\wh \phi}\,
\wh H_{NPQ}\,\wh\Gamma^{NPQ}\, \wh\Gamma_M\, \wh\e^{\un A} \ ,\nn\\
\d \wh \f &= \fr12 \wh {\bar \e}^{\un A}  \wh \c_{\un A}\ ,  &
\delta\wh\chi^{\un A} &=
-\fr14 ( \pa_M\wh\phi\, \wh\Gamma^M - \fr16 e^{\fr12\hat\phi}\, 
\wh H_{MNP}\, \wh\Gamma^{MNP} ) \wh\e^{\un A}\ , \nn \\
\d \wh{A}_M &= \fr1{2 \sq{2} } e^{-\fr14 \wh \f} \wh {\bar \e}^{ \un A}  \wh \G_M \wh \l_{\un A}\ , &
\delta \wh\lambda^{\un A} &=
\fr1{4\sqrt{2}}e^{\fr14\wh\phi}\, \wh F_{MN}\, \wh\Gamma^{MN}\wh \e^A -  \sq 2   \,
g\, e^{-\fr14\wh \phi}  \wh C^{\un{ A} \un {B} }  \wh\e_{\un B}\ , \nn \\
\d \wh L_{\uh \a}{}^{\uh a}  &= - \fr14  \wh {\bar \e}_\un A \wh \p^\uh a \wh L_{\uh \a }{}^{\un A}\ , &
\d \wh L_{\uh \a}{}^{\un  A} &= \fr14 \wh {\bar \e}^\un A \wh \p_\uh a \wh L_{\uh \a }{}^{\uh a}\ , \hspace{1.8 cm}
\d \wh \p^{\hat{\un a}}  =    \wh \G^M  \wh \e^{\un A} \wh P_M{}^\uh a{}_\un A\ , \nn \\
\d \wh B_{M N}  &= \wh  A_{[M} \d \wh A_{N]}  + \fr14 e^{-\fr12 \wh \f} \big(  2 \wh {\bar \e}^{\un A} \wh \G_{[M} \wh \p_{N] \un A}  + 2 \wh {\bar \p}_{[M}^{\un A} \wh \G_{N]} \wh \e_{\un A} + \wh {\bar \e}^{\un A} \wh \G_{M N } \wh \c_{\un A} - \wh {\bar \c}^{\un A} \wh \G_{MN} \wh \e_{\un A} \big) \ . \hspace{-8cm} &
\label{6DHyperSUSYtrans1}
\ea
The action also has a gauged $U(1)_R$ symmetry acting on the hypermultiplets. The embedding of this $U(1)_R$ within the $Sp(n,1)$ is described by the tensor $T_\uh \a{}^\uh \b$ which for now we leave general. However, we will later see that a consistent reduction is only possible for a particular $U(1)_R$ corresponding to a particular choice of embedding tensor. 
\subsection{Reduction of the Bosonic Sector including Hypermultiplets}
\label{ReductionBosonicHypermultiplets}
We shall now consider the dimensional reduction of the enlarged six-dimensional theory. We begin as before with the bosonic field equations; neglecting bifermionic terms, these read
\ba
\wh R_{M N} &= \fr14 \wh \pa_M \wh \f \wh \pa_N \wh \f + 2 \wh P_{M}^{\uh a \un A} \wh P_{N \uh  a \un A}  
+ \fr12 e^{\fr12 \wh \f} ( \wh F_{M R} \wh F_N{}^R - \fr18 \wh F_{RS} \wh F^{RS} \wh g_{MN})\nn \\
&\quad
 + \fr14 e^{ \wh \f } ( \wh H_{M RS} \wh H_{N}{}^{RS} - \fr16 \wh H_{RST} \wh H^{RST} \wh g_{MN} )  + g^2 \wh C^{\un{A} \un{B}} \wh C_{\un{A} \un{B}}  e^{-\fr12 \wh \f } \wh g_{MN}\ , \nn  \\
d \wh * d \wh \f &=  - \fr12 e^{\fr12 \wh \f}  \wh * \wh F_{\tw} \wedge \wh F_\tw    - e^{\wh \f} \wh * \wh H_\th \wedge \wh H_\th  + 4 g^2 \wh C^{\un{A} \un{B}} \wh C_{\un{A} \un{B}}  e^{-\fr12 \wh \f} \wh * 1\ , \nn \\
d ( e^{\fr12 \wh \f} \wh * \wh F_\tw ) &= e^{\wh \f} \wh * \wh H_\th \we \wh F_\tw \nn + 4 g \wh \x^{ \uh a \un A} \wh * \wh P_{\uh a \un A} \ ,  \hspace{2cm}
d ( e^{\wh \f} \wh * \wh H_\th ) = 0\ , \nn \\
\wh D \wh * \wh P_{\uh a}{}^{ \un A} &=  - 4 g^2 e^{ - \fr12 \wh \f} \wh C^{\un A \un B} \wh \x_{\uh a \un B}  \ , 
\ea
where the field strengths satisfy the Bianchi identities
\ba
d \wh H_\th &= \fr12 \wh F_\tw \we \wh F_\tw \ ,  & 
d \wh F_\tw &= 0\ . 
\ea
We then make an ansatz for a consistent set of four-dimensional fluctuations, similar to \eqref{BosonicAnsatz}, again obtained by trial and error:
\ba
\wh F_\tw &= \fr1{2g} \O_\tw - F_\tw\ , & 
 \nn \\
\wh H_\th &= H_\th - \fr{1}{2g} A_\on \we \O_\tw\ ,  &
\wh \f &= \vf - \f \ , \nn \\
\wh e^\a &= e^{\fr14 ( \vf + \f )  } e^\a\ , & 
\wh e^a &= e^{ - \fr14(  \vf + \f)  } e^a\ , \nn \\
\wh L_{\uh \a}{}^{\uh a} &= \til L_{\uh \a}{}^{\uh a}  \ ,  & 
\wh L_{\uh \a}{}^{\un A} &= \til L_{\uh \a}{}^{\un A} \ ,
\label{HyperAnsatz1}
\ea 
where the scalar functions $\til L_{\uh \a}{}^{\uh a}$ and $ \til L_{\uh \a}{}^{\un A} $ depend only on the uncompactified $x^\m$ directions. They satisfy the relations
\ba
\til L^\uh \a{}_{\un A} \til L_{\uh \a}{}^{\un B} &=  - \wh \d_\un A{}^\un B,  &
 \til L^\uh \a{}_{\uh a}  \til L_{\uh \a}{}^{\uh b} &=   \wh \d_\uh a{}^\uh b\ , \nn \\
\til L^\uh \a{}_{\un A} \til L_{\uh \a}{}^{\uh a} &= 0 \ , 
&  - \til L_{\uh \a}{}^{\un A} \til L^\uh \b{}_{\un A} + \til L_{\uh \a}{}^{\uh a}  \til L^\uh \b{}_{\uh a} &=   \wh \d_\uh \a{}^\uh  \b\ . 
\label{tilLIds}
\ea
and 
\ba
( \til L_\uh \a{}^\uh a )^* &=  - \til L^\uh \b{}_\uh a \wh \h_{\uh \b}{}^{\uh \a} &  ( \til L_\uh \a{}^\un A )^* &=  - \til L^\uh \b{}_\un A \wh \h_{\uh \b}{}^{\uh \a}
\ea
from these we can build the Maurer-Cartan forms
\ba
\til P_{ \m \uh a}{}^\un A &=  \til L^\uh \a{}_{\uh a} D_\m \til L_{\uh \a}{}^{\un A}\ , &
\til Q_{\m \un A}{}^\un B &= \til L^\uh \a{}_{\un A }  D_\m \til L_{\uh \a}{}^{\un B}\ ,  &
\til Q_{ \m \uh a}{}^\uh b &=  - \til L^\uh \a{}_{\uh a}  D_\m \til L_{\uh \a}{}^{\uh b}\ , 
\ea
where
\ba
D_\m \til L_\uh \a{}^\un A &= \pa_\m \til L_\uh \a{}^{\un A} - g  A_\m T_{\uh \a}{}^{\uh \b} \til L_{\uh \b}{}^{\un A}\ ,   & 
D_\m \til L_\uh \a{}^\uh a &= \pa_\m \til L_\uh \a{}^{\uh a} - g  A_\m T_{\uh \a}{}^{\uh \b} \til L_{\uh \b}{}^{\uh a}\ . 
\ea
We can also form the useful scalar functions
\ba
\til C_{\un A}{}^{\un B} &= \til L^{\uh  \a}{}_\un A T_{\uh \a}{}^{\uh \b} \til L_{\uh \b}{}^{\un B} \ ,& 
\til \x_{\uh a}{}^{\un A} &= \til L^{\uh \a}{}_\uh a T_{\uh \a}{}^{\uh \b} \til L_{\uh \b}{}^{\un A}\ . 
\ea
In this, we have left the ansatz for the hypermultiplet scalars very general, requiring only that they do not depend on the compactified directions. However, as we will see, requiring that the reduction be consistent places a large number of additional constraints on the scalar sector. Solving these constraints will require us to split the $\uh \a$, $\uh a$ and $\un A$ indices and will be solved later in Section \ref{SolutionConstraints}. Equation \eqref{HyperAnsatz1} therefore does not represent the final ansatz for the hypermultiplet scalars and will be further refined. 

The ansatz considered implies that
\ba
\wh * \wh F_\tw &= 4 g e^{\fr32 \f} * 1 - \fr1{8g^2} e^{-\fr12 \f} * F_\tw \we \O_\tw\ , &
\wh * 1 &= \fr1{8g^2} e^{\fr12 ( \f +\vf ) } * 1 \we \O_\tw\ , \nn \\
\wh * \wh H_\th &= \fr{1}{8g^2} e^{- \f} * H_\th \we \O_\tw - 4 g e^\f * A_\on\ , &
\wh * d \wh \f &= \fr1{8 g^2} * d ( \vf - \f ) \we \O_\tw\ , \nn  \\
\wh * \wh D \wh L_{\uh \a}{}^{\un A} &= \fr1{8 g^2} * D \til L_{\uh \a}{}^{\un A} \we \O_\tw   - g e^{ \vf + \f} T_\uh \a{}^\uh \b \til L_{\uh \b}{}^{\un A}   *_\tw A^{\text{mono}}_\on \we * 1 \ , \hspace{-2.5cm} \nn \\
\wh * \wh D \wh L_{\uh \a}{}^{\uh a} &= \fr1{8 g^2} * D \til L_{\uh \a}{}^{\uh a} \we \O_\tw   - g e^{ \vf + \f} T_\uh \a{}^\uh \b \til L_{\uh \b}{}^{\uh a}   *_\tw A^{\text{mono}}_\on \we * 1\ ,  \hspace{-2.5cm} 
\ea
Substituting this ansatz into the $\wh L_\uh \a{}^\un A$ field equation gives
\ba
D * (  \til L^{ \uh \a \uh a} D \til L_{\uh \a}{}^{ \un A} ) &=  - 4 g^2 e^{ \f}  \til C^{\un A \un B} \til \x^{\uh a}{}_{\un B} + 
8 g^3 e^{(\f +\vf) } \til \x^{\uh a \un A} d *_\tw A^{\text{mono}}_\on \we * 1 \nn \\ 
&\quad 
+ 8 g^4 e^{(\f +\vf) } ( \til C^{\uh a \uh b} \til \x_{\uh b}{}^{ \un A}  + \til C^{\un A \un B} \til \x^{\uh a}{}_{ \un B} )  *_\tw A^{\text{mono}}_\on \we A^{\text{mono}}_\on * 1\ , 
\ea
where $\til C^{\uh a \uh b}$ is defined analogously to $\til C^{\un A \un B} $. As $A^{\text{mono}}_\on$ is a function of the internal space, the reduction will only be consistent if $\til \x^{ \uh a \un A}  = 0$, so that the terms containing $A^{\text{mono}}_\on$ vanish. This represents the first of the constraints on the hypermultiplet ansatz and implies that 
\ba
D * \til P^{\uh a \un A} &=  0\ . 
\label{RedLEOM}
\ea
Substituting into the $\wh \f$ field equation, we find
\ba
d * d \f - d*d \vf &=  e^{- 2 \f} * H_\th  \we H_\th+ \fr12 e^{-\f } * F_\tw \we F_\tw  \nn \\
&\quad
 + 16 g^2 e^{2\vf} * A_\on \we A_\on - 4  g^2 e^\f (   \til C^{\un{A} \un{B}} \til C_{\un{A} \un{B}}  - 2   e^{2 \vf} ) * 1\ . 
\label{4DHyperphiEOM1}
\ea
Substituting into the $\wh R_{ab}$ field equation implies
\ba
d * d \f + d * d \vf  &=  e^{- 2\f} * H_\th  \we H_\th+ \fr12 e^{-\f } * F_\tw \we F_\tw \nn  \\
&\quad
 - 16 g^2 e^{2\vf} * A_\on \we A_\on - 4  g^2 e^\f (  \til C^{\un{A} \un{B}} \til C_{\un{A} \un{B}}   - 8 e^{\vf}   + 6  e^{2 \vf} ) * 1\ , 
\label{4DHyperphiEOM2}
\ea
where we have used $R_{m n} = 8 g^2 g_{m n}$. These equations can be rearranged to  give
\ba
d * d \f  &=  e^{ - 2 \f} * H_\th  \we H_\th+ \fr12 e^{- \f} * F_\tw \we F_\tw \nn \\
&\quad
 - 4 g^2 e^{ \f }  (   \til C^{\un{A} \un{B}} \til C_{\un{A} \un{B}}   - 2 + 2  ( 1- e^\vf)^2   ) * 1\ , \nn \\
d * d \vf &= - 16 g^2 e^{2 \vf} * A_\on \we A_\on + 16 g^2 e^{\f} e^{ \vf} ( 1 - e^{\vf} ) * 1\ .
\ea
As before, the $\wh R_{\a a}$ field equation gives just an identity $0 = 0$, but now substituting the ansatz into the $\wh R_{\a \b}$ field equation we find
\ba
R_{\a \b} &= \fr12 \pa_\a \f \pa_\b \f +  2 \til P_\a^{\uh a \un A} \til P_{\b \uh a \un A}  + \fr12 \pa_\a  \vf  \pa_\b \vf \nn \\
&\quad
 + \fr12 e^{- \f} ( F_{\a \g} F_\b{}^\g - \fr14 F_{\g\d} F^{\g \d} \h_{\a \b} )  
 + \fr14 e^{- 2 \f} ( H_{\a \g \d} H_{\b} {}^{\g \d} - \fr13 H_{\g \d \l} H^{\g \d \l} \h_{\a \b} )\nn \\
 &\quad
 +  2 g^2 e^\f (    \til C^{\un{A} \un{B}} \til C_{\un{A} \un{B}}   - 2 + 2 ( 1- e^\vf)^2   ) \h_{\a \b}  + 8 g^2 e^{2 \vf}  A_\a A_\b\ . 
\ea
The $ \wh H_\tw $ and $\wh F_\tw$ field equations  along with $\til \x^{\uh a \un A} = 0  $ together imply
\ba
d( e^{-2 \f} * H_\th) &= 0\ , &
d ( e^{ \f} * F_\tw )  &= e^{- 2 \f} * H_\th \we F_\tw + 16 g^2 e^{2\vf}  * A_\on\ .
\ea
Moreover, the Bianchi identities imply that
\ba
d H_\th &= \fr12 F_\tw \we F_\tw\ , &
d F_\tw &= 0\ . 
\ea
If the value $\til C^{\un A \un B} \til C_{\un A \un B} $ is not dependant on $\til L_{\uh \a}{}^{\un A}$, explaining why a potential term does not appear in \eqref{RedLEOM}, then these reduced field equations can all be derived from an action
\ba
\cL_B &= R * 1 - \fr12 * d \f \we d \f - \fr12 * d \vf \we d \vf    - 2 * \til P^{\uh a \un A} \we \til P_{\uh a \un A} \nn \\
&\quad
 - \fr12 e^{-2\f} * H_\th \we H_\th 
 - \fr12 e^{-\f} *F_\tw \we F_\tw - 8 g^2 e^{2\vf} * A_\on \we A_\on \nn \\
&\quad
 - 4 g^2 e^\f (  \til C^{\un{A} \un{B}} \til C_{\un{A} \un{B}}   - 2 + 2 ( 1- e^\vf)^2   ) *1\ , 
\ea
where $F_\tw = d A_\on$ and $H_\th = d B_\tw + \fr12 \o_\th$ and $d \o_\th = F_\tw \we F_\tw  $.
Finally, dualising $H_\th$ as in \eqref{dualH}  gives the bosonic part of the reduced action
\ba
\cL_B &= R * 1 - \fr12 * d \f \we d \f - \fr12 * d \vf \we d \vf    - 2 * \til P^{\uh a \un A} \we \til P_{\uh a \un A}  + \fr12 e^{ 2 \f} * d\s \we d \s \nn \\
&\quad
 - \fr12 e^{-\f} *F_\tw \we F_\tw - 8 g^2 e^{2\vf} * A_\on \we A_\on \nn  + \fr12 \s F_\tw \we F_\tw \\
&\quad
 - 4 g^2 e^\f (  \til C^{\un{A} \un{B}} \til C_{\un{A} \un{B}}   - 2 + 2 ( 1- e^\vf)^2   ) *1\ .
\ea

\subsection{Reduction of the Supersymmetry Transformations including Hypermultiplets}
\label{ReductionSusyHypermultiplets}
The symplectic Majorana and Weyl properties of the six-dimensional fermions cannot be carried over to the reduced theory because fermions simultaneously subject to four-dimensional symplectic Majorana and Weyl conditions identically vanish. This means that the $Sp(1)$ and $Sp(n)$ symmetries of the theory must be broken in the reduction. 
We therefore make an ansatz for the fermions equivalent to \eqref{FermionicAnsatz}, given by  
\ba
\wh \p_\a^1 &= e^{ - \fr18 ( \f + \vf)} \left[ \p_\a \ct \h + \fr1{2\sq 2} \g_\a ( \c + \z) \ct \h \right]\ ,&
\wh \l^1 &= e^{ - \fr18 ( \f + \vf)} \l \ct \h\ , \nn \\
\wh \p_a^1 &= - \fr1{ 2 \sq 2} e^{ - \fr18 ( \f + \vf)} ( \c + \z ) \ct \s_a \h\ , &
\wh \c^1 &= \fr1{\sq 2} e^{ - \fr18 ( \f + \vf)} ( \c - \z ) \ct \h\ , \nn \\
\wh \e^1 &=  e^{ \fr18 ( \f + \vf)} \e \ct \h\ .
\ea
The ansatz for $\wh \c^2$ can then be derived from this using \eqref{6DSymMaj} and similarly for the $2$ component of the other six-dimensional fermions. Here we note that since 
\ba
\wh \e^2&= i \wh C^{-1} \wh \G_0 ( \wh \e^1)^*\ , &&\text{and}&  \wh C &= C \ct \s^2 \ , 
\ea 
we have it that
\ba
 \s^3 \wh \e^1 &= \wh \e^1  &&\text{implies}& \s^3 \wh \e^2 &= - \wh \e^2\ , 
 \ea
and similarly the $1$ and $2$ components of the other six-dimensional fermions have opposite $\s^3$ and $\g^5$ chiralities. 

We make an ansatz for the reduction of the hypermultiplet fermions in a similar way to that considered above. First we split the index range of $\hat{\un a} $ into $\un a =1,...,n $ and $\un a^\pr = 1,...,n$ such that we write the $Sp(n)$ symplectic invariant as
\ba
\wh \e_{\hat{ \un a } \hat{\un b} } = \bordermatrix{  & \un a & \un a^{\pr}  \cr
 \un a & 0 & \id_n  \cr
 \un a^{\pr} & - \id_{n} & 0 } \ , 
\ea
so that $\wh \e_{\un a \un b} = \wh \e_{\un a^\pr \un b^\pr} = 0 $ and $\wh \e_{\un a \un b^\pr} = - \wh \e_{\un b^\pr \un a} = i_{\un a \un b^\pr} $ where  $i_{\un a \un a^\pr} = i_{\un a^\pr \un a}$ and $i_{\un a \un b^\pr} i^{\un b^\pr \un b} = \d_{\un a}{}^{\un b} $. 

We also split the $\uh \a$ index into $\un \a = 1, \ldots,  n + 1$ and $\un \a^\pr = 1, \ldots, n+1 $ such that we write the $Sp(n,1)$ symplectic invariant as
\ba
\wh \O_{\uh \a \uh \b } = \bordermatrix{ %
                         & \un \a= 1 & \un \a^\pr = 1 & \un \a \geq 2& \un \a^\pr \geq 2  \cr
 \un \a =   1                   & 0  & 1 & 0 & 0  \cr
\un \a^\pr = 1              & - 1 & 0 & 0 & 0 \cr 
\un \a \geq 2  & 0& 0 & 0 & - \id_n \cr
\un \a^{\pr} \geq 2  & 0 &  0&  \id_n & 0 }
 \ , 
\ea
so that $\wh \O_{\un \a \un \b} = \wh \O_{\un \a^\pr \un \b^\pr} = 0 $ and $\wh \O_{\un \a \un \b^\pr} = - \wh \O_{\un \b^\pr \un \a} = i_{\un \a \un \b^\pr} $ where  $i_{\un \a \un \a^\pr} = i_{\un \a^\pr \un \a}$ and $i_{\un \a \un \b^\pr} i^{\un \b^\pr \un \b} = \d_{\un \a}{}^{\un \b}\, $. Corresponding to this, the $U(2n,2)$ invariant is split as 
\ba
\wh \h_{\uh \a}{}^{ \uh \b } = \bordermatrix{ %
                         & \un \a= 1 & \un \a^\pr = 1 & \un \a \geq 2 & \un \a^\pr \geq 2  \cr
 \un \a =   1                   & 1  & 0 & 0 & 0  \cr
\un \a^\pr = 1              & 0 & 1 & 0 & 0 \cr 
\un \a \geq 2  & 0& 0 & -\id_n & 0  \cr
\un \a^{\pr} \geq 2  & 0 &  0&0& -\id_n }
 \ , 
\ea
so that $\wh \h_{\un \a}{}^{\un \b^\pr} = \wh \h_{\un \b^\pr }{}^{\un \a} = 0 $, $\wh \h_{\un \a}{}^{\un \b} = \h_{\un \a }{}^{\un \b}$ and $\wh \h_{\un \a^\pr}{}^{\un \b^\pr} = \h_{\un \a^\pr }{}^{\un \b^\pr}$ where 
\ba
\h_{\un \a}{}^{\un \b} &= \left( \begin{array}{cc}
 1&0\\
0 & - \id_n 
\end{array} \right) . 
\ea 
Using this index splitting we then make the ansatz
\ba
\wh \p^{\un a}  &= e^{ - \fr18 ( \f + \vf)} \p^{\un a}  \ct \h\ , \label{spinoransatz}
\ea
and note that since $\wh \G_7 \wh \p^{ \un a } = - \wh \p^{ \un a }$ , one finds $ \g_5 \p^{ \un a} = - \p^{\un a}$\ .

The four-dimensional fermions $\p^{\un a}$ describe $2n$ additional fermionic degrees of freedom arising from the six-dimensional hypermultiplets. This means that there cannot be a full set of $4n$ bosonic degrees of freedom in the reduced theory arising from the hypermultiplets as would appear if one turned on the full set of hypermultiplet scalars in \eqref{HyperAnsatz1}. It is for this reason that upon reduction we find the emergence of a large number of constraints on the hypermultiplet geometry emerge. 

Now consider the $\wh \l^1$ transformation 
\ba
\d \wh \l^{1} &=
\fr1{4\sqrt{2}} e^{\fr14\wh\phi}\, \wh F_{a b}\, \wh\Gamma^{ a b  } \wh \e^1 + \fr1{4\sqrt{2}} e^{\fr14\wh\phi}\, \wh F_{\a \b}\, \wh\Gamma^{ \a \b }\wh \e^1  \nn \\
&\quad -  \sq 2   \,
g\, e^{-\fr14\wh \phi} \wh C^{12  } \wh\e^{ 1} +  \sq 2  g\, e^{-\fr14\wh \phi} \wh C^{11  } \wh\e^{ 2} \ .\label{lambda1transf}
\ea
Imposing the ansatz $ \s^3 \wh \l^1 = \wh \l^1 $,  $ \s^3 \wh \e^1 = \wh \e^1 $ and $ \s^3 \wh \e^2 = - \wh \e^2 $ shows that the last term of \eqref{lambda1transf} must be zero and so we find another constraint on the four-dimensional scalar geometry, $ \til C^{11}  = 0\, $, as well as finding the resulting transformation of the four-dimensional spinor 
\ba
 \d \l &= - \fr1{4 \sq{2} } e^{- \fr12 \f} F_{\a \b} \g^{\a \b} \e + \sq{2} i g e^{\fr12 \f} e^{ \vf } -  \sq 2 g \til C^{12} e^\f  \e  \ . 
\ea
A similar consideration of the $\wh \l^2$ transformation shows that $ \til C^{22} = 0 $.
Considering the transformation of $\wh \p^{\un a}$ we find
\ba
\d \wh \p^{{\un a}} &=    \wh \G^\a  \wh \e^1 \wh P_\a{}^{{\un a} 2 } - \wh \G^\a  \wh \e^2 \wh P_\a{}^{{\un a} 1}\ . 
\ea
Using the same argument as above, we find $\til P_\m{}^{\un a 1} = 0$ and so 
\ba
\d \p^{{\un a}} &=   \g^\a  \e \til P_\a{}^{{\un a} 2 }\ . 
\ea
Similarly, the transformation of $\wh \p^{\un a^\pr}$ also implies $ \til P_\m{}^{\un a^\pr 2} = 0$. 
From the $\wh \p_\a$ transformation, we find the constraints $\til Q_\m{}^{ 1 1} = \til Q_\m{}^{ 2 2} = 0 $ and the transformation, 
\ba
\d \p_\a &= \na_\a \e -  \til Q_\a{}^{12} \e - i g e^\vf D_\a \r \e - \fr1{4} i e^{\f} \pa_\a \s \e\ . 
\ea
Considering the transformation of $\wh \p^1_a$ we find that in order to have $D_a \wh \e^1 = 0$ given that $(\na_a - i g A_{a}^{\text{mono}} )\h = 0$, this implies $\til C^1{}_1 = -i $. Similarly considering the variation of $\wh \p^2_a$, we find that $( \til C^2{}_2)^* = -i$ . 

Examining the transformations of $\til L_\uh \a{}^\un A$  and $\til L_\uh \a{}^\uh a$ we then find
\ba
\d  \til L_{\un \a}{}^{1} &= \fr14  {\bar \p}{}_\un a \e \til L_{\un \a }{}^{\un a}\ ,&
\d  \til L_{ \un \a}{}^{2} &= -  \fr14  {\bar \e} \p^\un a\til L_{\un \a \un a}\ , \nn \\
\d  \til L_{\un \a^\pr}{}^{1} &= \fr14  {\bar \p}{}_\un a \e \til L_{\un \a^\pr }{}^{\un a}\ ,&
\d  \til L_{ \un \a^\pr}{}^{2} &= -  \fr14  {\bar \e} \p^\un a \til L_{\un \a^\pr \un a}\ , \nn \\
\d  \til L_{\un \a}{}^{\un a}  &=  \fr14   {\bar \e}  \p^\un a \til L_{\un \a}{}^{ 1}\ ,&
\d  \til L_{\un \a}{}^{\un a^\pr}  &= \fr14  i^{\un a^\pr \un a} \bar \p{}_\un a \e \til L_{\un \a }{}^{2}\ ,\nn \\
\d  \til L_{\un \a^\pr}{}^{\un a}  &=  \fr14   {\bar \e}  \p^\un a \til L_{\un \a^\pr}{}^{1}\ ,&
\d  \til L_{\un \a^\pr}{}^{\un a^\pr}  &= \fr14  i^{\un a^\pr \un a} \bar \p{}_\un a \e \til L_{\un \a^\pr}{}^{2} \ ,
\ea
where $\bar \p_{\un a} = i ( \p^{\un a} )^\dagger \g_0$ and we have used $\h^T \s^2 \h = 0 $. The transformations of the other fields remain unaffected:
\ba
\d e_\m{}^\a &= - \fr14 ( \bar \e \g^{\a} \p_\m - {\bar \p}_\m  \g^\a \e ), &  \d A_\m &= - \fr1{2 \sq{2} } e^{\fr12 \f} ( \bar \e \g_\m \l - \bar \l \g_\m \e)\ ,\nn \\
\d \vf &= - \fr1{2\sq{2}} ( \bar \e \z + \bar \z \e)\ ,&
\d \z &= \fr1{2\sq{2}} \pa_\a \vf \g^\a \e + i g \sq{2} e^\vf D_\a \r \g^\a \e\ ,\nn \\
\d \f &= - \fr1{2\sq{2}} ( \bar \e \c + \bar \c \e)\ ,  & 
\d \c &= \fr1{2\sq{2}} \pa_\a \f \g^\a \e + \fr1{2\sq{2}} i e^{\f} \pa_\a \s \g^\a \e\ ,\nn \\
\d \s &=  \fr1{2\sq{2}} i e^{ - \f} ( \bar \e \c - \bar \c \e)\ ,& 
\d \r &=   \fr1{8\sq{2}g} i e^{ - \vf} ( \bar \e \z - \bar \z \e )\ .  \nn 
\ea

\subsection{Reduction of the Fermionic Sector including Hypermultiplets}
\label{ReductionFermionicHypermultiplets}

The reduction of the fermionic equations of motion proceeds much as before. We begin by considering the six-dimensional fermionic field equations 
\ba
\wh \G^{MNR} \wh D_N \wh \p_R^{\un A} &= - \fr14 \wh \G^N \wh \G^M \wh \c^{\un A} \wh \pa_N \wh \f -  \wh \G^N \wh \G^M \wh \p^{\hat {\un a} } \wh P_{N \uh a}{}^{\un A} \nn \\
&\quad
- \fr1{24} e^{\fr12 \wh \f} \wh H^{NRS} \big( \wh \G^{[M} \wh \G_{NRS} \wh \G^{T]} \wh \p^{\un A}_T   + \wh \G_{NRS} \wh \G^M \wh \c^{\un A} \big) \nn \\
&\quad 
+ \fr1{4 \sq{2} } e^{\fr14 \wh \f} \wh F_{NR} \wh \G^{NR} \wh \G^M \wh \l^{\un A} 
+ \sq{2} g e^{- \fr14 \wh \f} \wh C^{\un A}{}_{\un B} \wh \G^M \wh \l^{\un B}\ , \nn \\
\wh \G^M \wh D_M \wh \l^{ \un A} &= - \fr1{24} e^{\fr12 \wh \f} \wh H_{MNR} \wh \G^{MNR} \wh \l^{\un A} 
+ \fr1{4 \sq{2} } e^{\fr14 \wh \f} \wh F_{MN} \big( \wh \G^R \wh \G^{MN} \wh \p^{\un A}_R 
- \wh \G^{MN} \wh \c^{\un A} \big) \nn \\
&\quad
- 2 \sq 2 g e^{-\fr14 \wh \f} \wh \p^{\hat {\un a}} \wh  \x_{\uh a}{}^\un A
+  \sq{2} g e^{- \fr14 \wh \f} \wh C^{\un A}{}_{\un B} \big( \wh \G^M \wh \p^{\un B}_M + \wh \c^{\un B} \big)\ , \nn \\
\wh \G^M \wh D_M \wh \c^{\un A} &= - \fr14 \wh \G^N \wh \G^M \wh \p^{\un A}_N \wh \pa_M \wh \f 
+ \fr1{24} e^{\fr12 \wh \f} \wh H_{MNR} \big(  \wh \G^S \wh \G^{MNR} \wh\p^{\un A}_S + \wh \G^{MNR} \wh \c^{\un A} \big) \nn \\
&\quad
+ \fr1{4 \sq{2} } e^{\fr14 \wh \f} \wh F_{MN} \wh \G^{MN} \wh \l^{\un A}  
- \sq{2} g \wh  C^{\un A}{}_{\un B} e^{- \fr14 \wh \f} \wh \l^{\un B} \ , \nn \\
\wh \G^M \wh D_M \wh \p^{\hat {\un a} } &=  \wh \G^N \wh \G^M \wh \p^{\un A}_N \wh  P_{M }{}^\uh a{}_{\un A} + \fr1{24} e^{\fr12 \wh \f} \wh H_{MNR} \wh \G^{MNR} \wh \p^{\hat {\un a} }  
+  2 \sq 2 g e^{-\fr14 \wh \f} \wh \l^{\un A} \wh   \x^{ \uh a}{}_{\un A}  \ . 
\ea
Then substituting the ansatz and constraints into the $\wh \l$ field equation gives
\ba
\g^\a &(\na_\a +  \til Q_\a{}^{1}{}_{1} ) \l=\hfill\nn \\
& - \fr1{24} e^{- \f} H_{\a \b \g} \g^{\a\b\g}  \l + i g e^\vf A_\a \g^\a \l 
- \fr1{4 \sq{2} } e^{-\fr12 \f} F_{\a\b} \big( \g^\g \g^{\a \b} \p_{\g} - \sq 2 \g^{\a\b} \c \big) \nn \\
& + i g e^{\fr12 \f} ( e^\vf - 1) \big(  \sq 2 \g^\a\p_\a + 2 \c ) + i 4  g e^{\fr12 \f} e^\vf  \z\ . 
\ea
Considering the $\wh \p^{\un a} $ field equation, we find
\ba
& \g^\a ( \na_\a  \p^{\un a} + \til Q_{\a}{}^{\un a}{}_\un b \p^\un b)  =   -  \g^\a \g^\b \p_\a \til P_\b{}^{\un a 2}  
+ \fr1{24} e^{- \f} H_{\a \b \g} \g^{\a\b\g}  \p^{\un a}  
 - i g e^\vf A_\a \g^\a \p^{\un a}\ ,
\ea
but for consistency we are also forced to require that 
\ba
 \til Q_{\a}{}^{\un a}{}_{\un b^\pr} &= \til C^{\un a}{}_{\un b^\pr} = 0 && \text{and} & \G^a( \na_a  \wh \p^\un a - g A_a^{\text{mono}} \til C^{\un a}{}_\un b \wh \p^\un b ) = 0\ ,
 \ea
which implies $\til C_\un a{}^\un b = i \d_\un a{}^\un b$. Similarly the $\wh \p^{\un a^\pr}$ field equation implies that $(\til C_{\un a^\pr} {}^{\un b^\pr})^* = i \d_{\un a^\pr}{}^{\un b^\pr}\,$.
Substituting the ansatz into the $\wh \c$ field equation implies
\ba
\fr1{\sq 2 } \g^\a & ( \na_\a  + \til Q_\a{}^{1}{}_{1}) ( \c - \z ) =\nn \\
&- \fr14 \g^\a \g^\b \p_\a \pa_\b ( \vf - \f ) 
+ \fr1{24} e^{-\f} H_{\a\b\g} \big( \g^\d \g^{\a\b\g} \p_\d + \fr1{\sq 2 }  \g^{\a\b\g} \z + 3 \fr1{\sq 2 }  \g^{\a\b\g} \c \big) \nn \\
&
- i g e^\vf A_\a \big( \g^\b \g^\a \p_\b -\fr1{\sq 2 } \g^\a \c - 3 \fr1{\sq 2 } \g^\a \z \big) - \fr1{ 4 \sq 2} e^{- \fr12 \f} F_{\a\b} \g^{\a\b} \l \nn \\
&
+ i \sq 2 g e^{\fr12 \f} ( e^\vf + 1) \l\ .
\ea
As before, we must consider the $\wh \p_M$ field equations with the free index pointing in the compact and noncompact directions separately.  When the free index is in the $\alpha$ direction, this implies
\ba
\g^{\a \m \n} &( \na_\m  + \til Q_\m{}^{1}{}_{1}) \p_\n 
 - \fr1{4\sq 2}  \g^\b \g^\a ( \z + \c ) \pa_\b ( \f + \vf) = \nn \\
&
 - \fr1{4 \sq 2}  \g^\b \g^\a ( \c - \z ) \pa_\b ( \vf - \f) - \g^\b \g^\a \p^{\un a} \til P_{\b \un a 2}  
 + \fr1{4} e^{- \f} H^{\a \b \g} \g_\g \p_\b - \fr1{12 \sq 2} e^{-\f} H_{\b\g\d} \g^{\b\g\d} \g^\a \c \nn \\
&
  - i g e^\vf A_\g \big( \g^{\a\b\g} \p_\b + \sq 2 \g^\g \g^\a \z \big) 
 - \fr1{4 \sq 2} e^{- \fr12 \f } F_{\b\g} \g^{\b\g} \g^\a \l  
   + i \sq 2 g e^{\fr12 \f } ( e^\vf - 1 ) \g^\a \l\ ,
\ea
while the  $\wh \p_M$ field equation with the free index in the $a$ direction gives
\ba
 \g^{\m \n} &( \na_\m + \til Q_\m{}^{1}{}_{1} ) \p_\n + \fr1{\sq2} \g^\m ( \na_\m + \til Q_\m{}^{1}{}_{1}) ( \c + \z) \nn \\ 
 - \fr14 &\pa_\n ( \f + \vf)  \g^\m \g^\n \p_\m + \fr1{4\sq 2} \pa_\m ( \f + \vf) \g^\m ( \z + \c) =\nn \\
&  \fr1{4\sq2} \g^\a \pa_\a ( \vf - \f ) ( \c - \z ) + \g^\a \p^{\un a} \til P_{\a \un a 2} 
- \fr1{24} e^{-\f} H_{\a\b\g} \big( \g^{\a \b\g\d} \p_\d - \fr1{\sq2} \g^{\a\b\g} \c + \fr1{\sq2} \g^{\a\b\g} \z \big) \nn \\ 
&+ i g e^\vf A_\a \big( \p^\a  + \fr1{\sq2} \g^\a \c - \fr1{\sq2} \g^\a \z \big)  
- \fr1{4 \sq 2} e^{- \fr12 \f } F_{\a \b} \g^{\a\b} \l   
 - i \sq 2 g e^{\fr12 \f} ( e^\vf +1) \l\ .
\ea

\subsection{Solution of the Hypermultiplet Constraints}
\label{SolutionConstraints}
In carrying out this reduction, we have found a large number of constraints on the hypermultiplet scalars. These read
\ba
 \til P_\m{}^{\un a^\pr 2} &= 0\ ,&
 \til  P_\m{}^{\un a 1} &= 0\ ,&
\til \x^{ \uh a \un A}  &= 0\ ,\nn \\
\til Q_\m{}^{ 1 1} &= 0\ , &  
\til Q_\m{}^{ 2 2} &= 0\ ,& 
\til Q_{\m}{}^{\un a}{}_{\un b^\pr} &= 0\ ,& 
\til Q_{\m}{}^{\un a^\pr}{}_{\un b} &= 0\ ,\nn \\
\til  C^{11}  &= 0\ , &  
 \til C^{22} &= 0\ ,& 
\til C^1{}_1 &= -i\ ,&
\til C^2{}_2 &= i\ , \nn \\
\til C^{\un a}{}_{\un b^\pr} &= 0\ ,& 
 \til C^{\un a^\pr}{}_{\un b} &= 0\ ,&
 \til C_\un a{}^\un b &= i \d_\un a{}^\un b\ , & 
\til C_{\un a^\pr}{}^{\un b^\pr} &=  - i \d_{\un a^\pr}{}^{\un b^\pr}\ .
\ea
Using the identity \eqref{LIds}, we can rearrange the $\til C$ and $\til \x$ constraints to give
\ba
T_{\uh \a}{}^{\uh \b} \til L_{\uh \b 1} &= - i \til L_{\uh \a 1}\ ,& 
T_{\uh \a}{}^{\uh \b} \til L_{\uh \b 2} &=  i \til L_{\uh \a 2}\ , \nn \\ 
T_{\uh \a}{}^{\uh \b} \til L_{\uh \b \un a} &= - i \til L_{\uh \a \un a}\ ,& 
T_{\uh \a}{}^{\uh \b} \til L_{\uh \b \un a^\pr} &=  i \til L_{\uh \a \un a^\pr}\ .
\label{Tconst}
\ea
Clearly, one solution is to have all the hypermultiplet fluctuations vanish, in which case we revert to the case studied before. However if we wish to consider a reduction where some hypermultiplet fluctuations are turned on, then we require also that the conditions $  \til P_\m{}^{\un a^\pr 1}  \neq 0$ and $ \til  P_\m{}^{\un a 2} \neq 0 $ be satisfied. Expanding out the constraints on $\til P_\m{}^{\uh a \un A}$ and using the constraint $\til \x^{ \uh a \un A}  = 0$ then gives
\ba
\til P_{\m \un a 1} &= i^{\un \a \un \a^\pr }(  \til L_{\un \a^\pr 1 } \pa_\m \til L_{ \un \a \un a} -  \til L_{\un \a 1 } \pa_\m \til  L_{ \un \a^\pr \un a} ) = 0\quad\;\text{but}& 
\til P_{\m \un a 2} &= i^{\un \a \un \a^\pr }(  \til L_{\un \a^\pr 2 } \pa_\m \til L_{ \un \a \un a} - \til L_{\un \a 2 } \pa_\m \til L_{ \un \a^\pr \un a} ) \neq 0\ ,\nn \\
\til P_{\m \un a^\pr 1} &= i^{\un \a \un \a^\pr }(  \til L_{\un \a^\pr 1 } \pa_\m \til L_{ \un \a \un a^\pr} - \til L_{\un \a 1 } \pa_\m \til L_{ \un \a^\pr \un a^\pr} ) = 0\quad\text{but} & 
\til P_{\m \un a^\pr 2} &= i^{\un \a \un \a^\pr }(  \til L_{\un \a^\pr 2 } \pa_\m \til L_{ \un \a \un a^\pr} - \til L_{\un \a 2 } \pa_\m \til L_{ \un \a^\pr \un a^\pr} ) \neq 0\ .
\label{Pconstraints}
\ea
Keeping in mind that the hypermultiplet scalars must also solve \eqref{LIds},  the constraints \eqref{Pconstraints} are solved by either 
\ba
\til L_{\un \a \un a}\ ,  \til L_{\un \a 1}\ , \til L_{\un \a^\pr  \un a^\pr}\ ,  \til L_{\un \a^\pr 2} & \neq 0 &
& \text{and} &
\til L_{\un \a^\pr \un a}\ ,  \til L_{\un \a^\pr 1}\ , \til L_{\un \a  \un a^\pr}\ ,  \til L_{\un \a 2}  &=  0
\ea
or 
\ba
\til L_{\un \a^\pr \un a}\ ,  \til L_{\un \a^\pr 1}\ , \til L_{\un \a  \un a^\pr}\ ,  \til L_{\un \a 2}  & \neq 0 &
& \text{and} &
\til L_{\un \a \un a}\ ,  \til L_{\un \a 1}\ , \til L_{\un \a^\pr  \un a^\pr}\ ,  \til L_{\un \a^\pr 2} &= 0\ .
\ea
Either choice is equivalent and here we chose the former. Then, substituting into \eqref{Tconst}, we find
\ba
T_{\un \a}{}^{\un \b} \til L_{\un \b 1} &= - i \til L_{\un \a 1}\ ,& 
T_{\un \a^\pr}{}^{\un \b^\pr} \til L_{\un \b^\pr 2} &=  i \til L_{\un \a^\pr 2}\ ,\nn \\ 
T_{\un \a}{}^{\un \b} \til L_{\un \b \un a} &= - i \til L_{\un \a \un a}\ ,& 
T_{\un \a^\pr}{}^{\un \b^\pr} \til L_{\un \b^\pr \un a^\pr} &=  i \til L_{\un \a^\pr \un a^\pr}\ ,
\ea
which is solved by 
\ba
T_{ \un \a}{}^{\un \b} &= - i \d_{\un \a}{}^{\un \b} \ ,&
 T_{ \un \a^\pr}{}^{\un \b^\pr} &=  i \d_{\un \a^\pr}{}^{\un \b^\pr} \ .
\ea
This fixes the gauging that was left general in the six-dimensional theory as mentioned in Section \ref{ReductionBosonicHypermultiplets} and solves the full set of hypermultiplet constraints. 

As we now know which of the six-dimensional hypermultiplet scalars must vanish and which can remain, we now refine our ansatz to 
\ba
\til L_{\un \a \un a} &= L_{\un \a \un a}\ , &
\til L^{\un \a \un a} &= \bar L^{\un \b \un a}  , \nn \\
\til L_{\un \a 1} &=  L_\un \a \ ,& 
\til L^{\un \a 1}&= \bar L^\un \b \ ,\nn \\
\til L_{\un \a^\pr \un a} &=  \til L^{\un \a^\pr \un a} = \til L_{\un \a^\pr 1} = \til L^{\un \a^\pr 1} =  0 \ .
\label{HyperAnsatz2}
\ea
These four-dimensional scalars then satisfy
\ba
\bar L^\un \a  L_\un \a &= 1\ , &
\bar L^{\un \a \un b}L_{\un \b \un a}  &= - \d_\un a{}^{\un b}\ , \nn \\
L_\un \a \bar L^\un \b - L_{\un \a \un a} \bar L^{\un \b  \un a } &= \d_\un \a{}^\un \b \ ,&
 \bar L^{\un \a} \h_{\un \a}{}^{\un \b} L_{\un \b \un a } &= 0\ ,
\ea
and
\ba
(L_{\un \a \un a})^* &= \bar L^{\un \b \un a} \h_{\un \b}{}^{\un \a} & (L_{\un \a})^* &= \bar L^{\un \b} \h_{\un \b}{}^{\un \a}
\ea
which implies that the surviving hypermultiplet scalars describe the coset 
\ba
\fr{ SU(n , 1)}{SU(n) \ct U(1)}\ ,
\ea
the dimension of which is $2n$. This agrees with the $2n$ fermionic degrees of freedom described by $\p^\un a\,$. 
We can then use these  scalars to define the four-dimensional Maurer-Cartan forms that will appear in the four-dimensional action
\ba
Q_\m &= \bar L^\un \a \pa_\m L_\un \a\ , &
 Q_{ \m}{}^\un a{}_\un b &= - \bar L^{\un \a \un a} \pa_\m L_{\un \a \un b}\ , \nn \\
P_{\m \un a} &= \bar L^\un \a \pa_\m L_{\un \a \un a}\ ,&
\bar P_{\m}{}^{ \un a} &=  L_\un \a \pa_\m \bar L^{\un \a \un a}\ .
\ea
\subsection{The Reduced Hypermultiplet Coupled  Action}
Substituting the modified ansatz \eqref{HyperAnsatz2} into the reduced field equations and rearranging gives 
\ba
\g^{\m \n \r} D_\n \p_\r &=
 \fr1{2\sq 2}  \g^\n \g^\m \c \pa_\n  \f +  \fr1{2\sq 2} \g^\n \g^\m \z \pa_\n  \vf  -   \g^\n \g^\m \p^{\un a} P_{\n \un a}  \  \nn \\
 &\quad
 + \fr1{4} e^{- \f} H^{\m \n \r} \g_\r \p_\n - \fr1{12 \sq 2} e^{-\f} H_{\r \s \t} \g^{\r\s\t} \g^\m \c \nn \\ 
 &\quad
  - i g e^\vf A_\r \big( \g^{\m\n\r} \p_\n + \sq 2 \g^\r \g^\m \z \big) 
- \fr1{4 \sq 2} e^{- \fr12 \f } F_{\r \s} \g^{\r \s} \g^\m \l  
 + i \sq 2 g e^{\fr12 \f } ( e^\vf - 1 ) \g^\m \l\ , \nn \\
\g^\a D_\a  \l &= 
- \fr1{24} e^{- \f} H_{\a \b \g} \g^{\a\b\g}  \l + i g e^\vf A_\a \g^\a \l 
- \fr1{4 \sq{2} } e^{-\fr12 \f} F_{\a\b} \big( \g^\g \g^{\a \b} \p_{\g} - \sq 2 \g^{\a\b} \c \big) \nn \\
&\quad
+ i g e^{\fr12 \f} ( e^\vf - 1) \big(  \sq 2 \g^\a\p_\a + 2 \c ) + i 4  g e^{\fr12 \f} e^\vf  \z\ ,\nn \\
\g^\m D_\m \z &= 
\fr1{2 \sq 2} \g^\m \g^\n \p_\m \pa_\n \vf
- \fr1{24} e^{-\f} H_{\m\n\r} \g^{\m \n\r} \z
+ i  g e^\vf A_\m (  \sq 2 \g^\n \g^\m \p_\n  - 3 \g^\m \z ) 
- i4 g e^{\fr12 \f } e^\vf \l\ ,\nn \\
\g^\m D_\m \c &= \fr1{2\sq2} \g^\m \g^\n \p_\m \pa_\n \f 
+ \fr1{24} e^{-\f} H_{\m\n\r} \big( \sq 2 \g^\s \g^{\m\n\r} \p_\s + 3 \g^{\m \n\r} \c \big) 
+ i g e^\vf A_\m \g^\m \c \nn \\
&\quad
- \fr1{4} e^{-\fr12 \f} F_{\m \n} \g^{\m \n} \l - i   2 e^{\fr12 \f } ( e^\vf - 1 )\ ,\nn \\
\g^\a D_\a  \p^{\un a}  &=   - \g^\a \g^\b \p_\a \bar P_\b{}^{\un a}  
+ \fr1{24} e^{- \f} H_{\a \b \g} \g^{\a\b\g}  \p^{\un a} 
 - i g e^\vf A_\a \g^\a \p^{\un a}\ ,
\ea
where 
\ba
D_\m \c &= \na_\m \c + i g A_\m \c + Q_\m \c\ ,
\ea
and similarly for all spinors except $\p^\un a$ for which 
\ba
D_\m \p^\un a &= \na_\m \p^\un a  + i g A_\m \p^\un a  + Q_\m{}^\un a{}_\un b \p^\un b\ . 
\ea
Integrating these field equations and combining with the bosonic part of the action as derived in Section  \ref{ReductionBosonicHypermultiplets}, we find that the full reduced hypermultiplet-coupled action is given by 
\ba
S &= \int d^4 x e \bls R - \fr12 \pa_\m \f \pa^\m \f - \fr12 \pa_\m \vf \pa^\m \vf - \fr12 e^{2\f} \pa_\m \s \pa^\m \s - 8 g^2 e^{2\vf} D_\m \r D^\m \r  - 4 P_{\m \un a} \bar P^{\m \un a} \nn \\
&\quad 
  - \fr14 e^{-\f} F_{\m \n} F^{\m \n}  - \fr18 \s \e^{\m\n\r\s} F_{\m \n} F_{\r \s}   - 4 e^\f g^2 ( 1-e^\vf)^2  
\nn \\
&\quad
+ \bar \p_\m \g^{\m \n \r} D_{\n} \p_{\r} + \bar \l \g^\m D_\m \l  +  \bar \c \g^\m D_\m \c +  \bar \z \g^\m D_\m \z \nn +  \bar \p{}_{\un a} \g^\m D_\m \p^{\un a}
 \nn \\
 &\quad
 - \fr1{2 \sq{2} } \big( \bar \p_\m \g^\n \g^\m \c + \bar \c \g^\m \g^\n \p_\m \big) \pa_\n \f 
- \fr1{2 \sq{2} } \big( \bar \p_\m \g^\n \g^\m \z + \bar \z \g^\m \g^\n \p_\m \big) \pa_\n \vf
\nn \\
&\quad
 +  \big( \bar \p_\m \g^\n \g^\m \p^{\un a} P_{\n \un a}  +  \bar \p_{\un a} \g^\m \g^\n \p_\m \bar P_\n{}^{\un a} \big) 
 \nn \\ 
 &\quad
+ \fr1{4} i e^{\f} \pa_{\m} \s \big ( \bar \p_\n \g^{\m \n \r} \p_\r + \sq 2 \bar \p_\n \g^{\m} \g^\n \c - \sq 2 \bar \c \g^\n \g^{\m} \p_\n  + 3 \bar \c \g^{\m} \c  -  \bar \z \g^{\m} \z +  \bar \l \g^{\m} \l  +  \bar \p_{\un a} \g^{\m} \p^{\un a} \big)
 \nn \\
 &\quad
+ i g e^\vf D_\m \r \big ( \bar \p_\n \g^{\m \n \r} \p_\r + \sq 2 \bar \p_\n \g^\m \g^\n \z - \sq 2 \bar \z \g^\n \g^\m \p_\n  + 3 \bar \z \g^\m \z  -  \bar \c \g^\m \c - \bar \l \g^\m \l    +  \bar \p_{\un a} \g^{\m} \p^{\un a} \big ) 
\nn \\
&\quad
+ \fr1{8 } e^{-\fr12 \f } F_{\m \n} \big( \sq 2 \bar \p_\r \g^{\m \n } \g^\r \l + \sq 2 \bar \l \g^\r \g^{\m \n} \p_\r + 2 \bar \c \g^{\m \n} \l - 2 \bar \l \g^{\m \n } \c  \big) 
\nn \\
&\quad 
- i g e^{\fr12 \f } ( e^\vf - 1) \big(\sq 2 \bar \p_\m \g^\m \l + \sq 2 \bar \l \g^\m \p_\m +  2 \bar \l \c - 2 \bar \c \l\big) 
- i  4  g e^{\fr12 \f } e^\vf ( \bar \l \z - \bar \z \l ) \brs\ ,
\ea 
which is supersymmetric under the transformations
\ba
\d e_\m{}^\a &= - \fr14 ( \bar \e \g^{\a} \p_\m - {\bar \p}_\m  \g^\a \e )\ ,&
\d \p_\a &= D_\a \e   - i g e^\vf D_\a \r \e - \fr1{4} i e^{\f} \pa_\a \s \e\ , \nn \\
\d A_\m &= - \fr1{2 \sq{2} } e^{\fr12 \f} ( \bar \e \g_\m \l - \bar \l \g_\m \e)\ ,&
\d \l &= - \fr1{4 \sq{2} } e^{- \fr12 \f} F_{\a \b} \g^{\a \b} \e + \sq{2} i g e^{ \fr12 \f } ( e^{\vf}  - 1)\ , \nn \\
\d \vf &= - \fr1{2\sq{2}} ( \bar \e \z + \bar \z \e)\ ,&
\d \z &= \fr1{2\sq{2}} \pa_\a \vf \g^\a \e + i g \sq{2} e^\vf D_\a \r \g^\a \e\ , \nn \\
\d \f &= - \fr1{2\sq{2}} ( \bar \e \c + \bar \c \e),  & 
\d \c &= \fr1{2\sq{2}} \pa_\a \f \g^\a \e + \fr1{2\sq{2}} i e^{\f} \pa_\a \s \g^\a \e\ , \nn \\
\d \s &=  \fr1{2\sq{2}} i e^{ - \f} ( \bar \e \c - \bar \c \e)\ ,& 
\d \r &=   \fr1{8\sq{2}g} i e^{ - \vf} ( \bar \e \z - \bar \z \e )\ , \nn \\
\d  L_{ \un \a} &=  \fr14  {\bar \e} \p^\un a L_{\un \a \un a}\ ,& 
\d  L_{\un \a \un a}   &=  \fr14   \bar \p{}_\un a \e L_{\un \a}\ , \nn \\
\d  \bar L^{\un \a}   &=  \fr14   \bar \p{}_\un a \e \bar L^{\un \a \un a}\ , &
\d  \bar L^{ \un \a \un a} &=  \fr14  {\bar \e} \p^\un a \bar L^{\un \a} \nn\ ,\\
\d \p^{{\un a}} &=   \g^\a  \e \bar P_\a{}^{{\un a} }\ .
\label{4DHyperSusy}
\ea
The spinors $\p^\un a$ appearing in this action are chiral fermions charged under a complex representation of $SU(n)$, giving a genuinely chiral theory in 
four dimensions.

\section{Gauging Hypermultiplet Symmetries}
\label{GaugingHyperSymmetries}

A further extension of the six-dimensional model is also possible in which additional symmetries of the hypermultiplets are gauged. To do this, we introduce the additional six-dimensional vector multiplet $( \wh A^I_M, \wh \l^I )$ such that the  action becomes 
\ba
S &= \int d^6 x \wh e \bls \wh R - \fr14 \wh \pa_M \wh \f \wh \pa^M \wh \f - 2 \wh P_{M \uh a \un A}  \wh P^{M \uh a \un A}  -   \fr1{12} e^{ \wh \f } \wh H_{MNR} \wh H^{MNR}  \nn \\
&\quad
- \fr14 e^{\fr12 \wh \f} \wh F_{MN} \wh F^{MN}  - \fr14 e^{\fr12 \wh \f} \wh F_{I MN} \wh F^{I MN}-   4 g^2 \wh C^{ \un {A} \un { B} }  \wh C_{\un{A} \un{B} } e^{-\fr12 \wh \f}  -   4 \tilde g^2 \wh C^{ I \un {A} \un { B} }  \wh C_{I \un{A} \un{B} } e^{-\fr12 \wh \f} \nn \\
&\quad
 + \fr12 \wh{\bar\p}^{\un A}_M \wh \G^{MNR} \wh D_N \wh \p_{R {\un A}} + \fr12 \wh{\bar \c}^{{\un A}} \wh \G^M \wh D_M \wh \c_{{\un A}}  + \fr12 \wh{\bar \l}^{{\un A}} \wh \G^M \wh D_M \wh \l_{{\un A}} + \fr12 \wh{\bar \l}^{I {\un A}} \wh \G^M \wh D_M \wh \l_{I {\un A}} + \fr12 \wh{\bar \p}{}^{\hat{\un a}}  \wh \G^M \wh D_M \p_{\hat{\un a}}   \nn \\ 
&\quad
+ \fr14 \wh{ \bar \c}^{\un A} \wh \G^N \wh \G^M \wh \p_{N{\un A}}  \wh \pa_{M} \wh \f  -   \wh{ \bar \p}^{\hat{\un a}} \wh \G^N \wh \G^M \wh \p_{N}^{{\un A}}   P_{ M \uh a \un A} \nn \\
&\quad
 + \fr1{48} e^{\fr12 \wh \f } \wh H_{MNR} \big( \wh{\bar\p}{}^{S{\un A}} \wh \G_{[S} \wh \G^{MNR} \wh \G_{T]} \wh \p{}^{T}_{\un A}  + 2 \wh{\bar\p}_S^{\un A} \wh \G^{MNR} \wh \G^S \wh \c_{\un A} - \wh{\bar \c}^{\un A} \wh \G^{MNR} \wh \c_{\un A} \nn \\
 &\quad
 + \wh{\bar\l}^{\un A} \wh \G^{MNR} \wh \l_{\un A}   + \wh{\bar\l}^{I \un A} \wh \G^{MNR} \wh \l_{I \un A} -  \wh{ \bar \p}{}^{\hat{\un a}} \G^{ MNR } \wh \p_{\hat{\un a}}\big) \nn \\
&\quad
 - \fr{1}{4 \sq{2} } e^{\fr14 \wh \f} \wh F_{M N} \big( \wh{\bar \p}_R^{\un A} \wh \G^{M N} \wh \G^R \wh \l_{\un A} + \wh{\bar \c}^{\un A} \wh \G^{MN} \wh \l_{\un A} \big)  - \fr{1}{4 \sq{2} } e^{\fr14 \wh \f} \wh F_{M N }^I \big( \wh{\bar \p}_R^{\un A} \wh \G^{M N} \wh \G^R \wh \l_{I \un A} + \wh{\bar \c}^{\un A} \wh \G^{MN} \wh \l_{I \un A} \big) \nn \\
&\quad
-  2 \sq 2 g e^{ - \fr14 \wh \f}  \wh {\bar \p}{}^{\hat{\un a}} \wh \l^{ \un A} \wh \x_{\uh a \un A}  - \sq{2}  g e^{- \fr14 \wh \f} \wh C_{\un{A} \un{B}} \big(  \wh{\bar\l}{}^{\un A} \wh \G^M \wh \p_M^{\un B} + \wh{\bar\l}{}^{\un A} \wh \c^{\un B} \big) \nn \\
&\quad
-  2 \sq 2 \tilde g e^{ - \fr14 \wh \f}  \wh {\bar \p}{}^{\hat{\un a}} \wh \l^{I  \un A} \wh \x_{I \uh a \un A}  - \sq{2}  \tilde g e^{- \fr14 \wh \f} \wh C_{I \un{A} \un{B}} \big(  \wh{\bar\l}{}^{I \un A} \wh \G^M \wh \p_M^{\un B} + \wh{\bar\l}{}^{I \un A} \wh \c^{\un B} \big) 
\brs\ ,
\label{6DGaugedHyperAction}
\ea 
where now 
\ba
\wh H_{MNR} &= 3 \wh \pa_{[M} \wh B_{N R]} + \fr32 \wh F_{[M N} \wh A_{R]} + \fr32 \wh \O_{M N R}\ ,\nn \\
\wh \O_{MNR} &=  \wh F^I_{[M N} \wh A_{R] I } - \fr13 f_{I J K} \wh A^I_{ [M  } \wh A_{N }^J \wh A_{R ]}^K\ ,\nn \\
\wh D_M \wh L_\uh \a{}^\un A &= \wh \pa_M \wh L_\uh \a{}^{\un A} + g \wh A_M T_{\uh \a}{}^{\uh \b} \wh L_{\uh \b}{}^{\un A}  + \tilde g \wh A_{ I M}  T^I{}_{\uh \a}{}^{\uh \b} \wh L_{\uh \b}{}^{\un A} \ ,\nn \\
\wh D_M \wh L_\uh \a{}^\uh a &= \wh \pa_M \wh L_\uh \a{}^{\uh a} + g \wh A_M T_{\uh \a}{}^{\uh \b} \wh L_{\uh \b}{}^{\uh a}+ \tilde g \wh A_{I M} T^I{}_{\uh \a}{}^{\uh \b} \wh L_{\uh \b}{}^{\uh a}\ , \nn \\
\wh \x_{I \uh a}{}^{\un A} &= \wh L^{\uh \a}{}_\uh a T_{I \uh \a}{}^{\uh \b} \wh L_{\uh \b}{}^{\un A}\ ,  \hspace{1cm}
\wh C_{I \un A}{}^{\un B}  = \wh L^{\uh  \a}{}_\un A T_{I \uh \a}{}^{\uh \b} \wh L_{\uh \b}{}^{\un B}\ ,
\ea
with all other definitions unchanged.
The additional fields transform under supersymmetry as
\ba 
\d \wh{A}^I_M &= \fr1{2 \sq{2} } e^{-\fr14 \wh \f} \wh {\bar \e}^{ \un A}  \wh \G_M \wh \l^I_{\un A}\ , &
\delta \wh\lambda^{I \un A} &=
\fr1{4\sqrt{2}}e^{\fr14\wh\phi}\, \wh F^I_{MN}\, \wh\Gamma^{MN} \wh \e^{\un A}  -  \sq 2   \,
\tilde g\, e^{-\fr14\wh \phi} \wh C^{I \un{ A} \un{B} } \wh\e_{\un B}\ .
\ea
and the transformation of $B_{MN}$ is modified to
\ba
\d \wh B_{M N} &= \wh  A_{[M} \d \wh A_{N]} + \wh  A^I_{[M} \d \wh A_{N] I}  + \fr14 e^{-\fr12 \wh \f} \big( 2 \wh {\bar \e}^{\un A} \wh \G_{[M} \wh \p_{N] \un A}  + 2 \wh {\bar \p}_{[M}^{\un A} \wh \G_{N]} \wh \e_{\un A} + \wh {\bar \e}^{\un A} \wh \G_{M N } \wh \c_{\un A} - \wh {\bar \c}^{\un A} \wh \G_{MN} \wh \e_{\un A} \big) \ .
\ea
The supersymmetry transformations of all other six-dimensional fields remain as shown in \eqref{6DHyperSUSYtrans1}. As before, we carry out the reduction by making the bosonic ansatz
\ba
\wh F_\tw &= \fr1{2g} \O_\tw - F_\tw\ ,& 
\wh F^I_\tw &= - F^I_\tw\ ,\nn \\
\wh H_\th &= H_\th - \fr{1}{2g} A_\on \we \O_\tw\ ,&
\wh \f &= \vf - \f \ ,\nn \\
\wh e^\a &= e^{\fr14 ( \vf + \f )  } e^\a\ ,& 
\wh e^a &= e^{ - \fr14(  \vf + \f)  } e^a\ , \nn \\
\wh L_{\un \a \un a} &= L_{\un \a \un a}\ , &
\wh L^{\un \a \un a} &= \bar L^{\un \a \un a}\ , \nn \\
\wh L_{\un \a 1} &=  L_\un \a\ ,&
\wh L^{\un \a 1} &= \bar L^\un \a \nn \ ,\\
\wh L_{\un \a^\pr \un a} &=  \wh L^{\un \a^\pr \un a} = \wh L_{\un \a^\pr 1} = \wh L^{\un \a^\pr 1} =  0\ , 
\label{GaugedHyperBosonicAnsatz}
\ea
and the fermionic ansatz 
\ba 
\wh \p_\a^1 &= e^{ - \fr18 ( \f + \vf)} \left[ \p_\a \ct \h + \fr1{2\sq 2} \g_\a ( \c + \z) \ct \h \right]\ , &
\wh \c^1 &= \fr1{\sq 2} e^{ - \fr18 ( \f + \vf)} ( \c - \z ) \ct \h\ ,\nn \\
\wh \p_a^1 &= - \fr1{ 2 \sq 2} e^{ - \fr18 ( \f + \vf)} ( \c + \z ) \ct \s_a \h\ , &
\wh \p^{\un a}  &= e^{ - \fr18 ( \f + \vf)} \p^{\un a}  \ct \h\ ,\nn \\
\wh \l^1 &= e^{ - \fr18 ( \f + \vf)} \l \ct \h\ , &
\wh \l^{I 1} &= e^{ - \fr18 ( \f + \vf)} \l \ct \h\ ,\nn \\
\wh \e^1 &=  e^{ \fr18 ( \f + \vf)} \e \ct \h\ .
\label{GaugedHyperFermionicAnsatz}
\ea
The reduction then proceeds exactly as described in Section \ref{HypermultipletCouplings} except that upon considering the $\wh \p_\a^1$ supersymmetry transformation and the $\wh\p^\un a$ field equation, we find the additional constraints
\ba
\wh A_{ \m I } \wh C^{I 1}{}_2 = \wh A_{ \m I }  \wh C^{I 2}{}_{1} = \wh A_{ \m I }  \wh C^{I \un a}{}_{\un a^\pr} = \wh A_{ \m I }  \wh C^{I \un a^\pr}{}_{\un a} = 0\ .
\ea
Using our ansatz \eqref{HyperAnsatz2}, we find that these constraints are solved by
\ba
\wh A_{ \m I } T^I{}_{ \un \a}{}^{\un \b^\pr} = \wh A_{ \m I } T^I{}_{\un \a^\pr }{}^{\un \b} = 0\ ,
\label{GaugeBreakingConditions}
\ea
which means that the reduction breaks the six-dimensional gauging of some subgroup of $Sp(n,1)$ to the part of that subgroup that lies within $SU(n,1)$. 

The reduced field equations found in this way can be integrated to give the action 
\ba
S &= \int d^4 x e \bls R - \fr12 \pa_\m \f \pa^\m \f - \fr12 \pa_\m \vf \pa^\m \vf - \fr12 e^{2\f} \pa_\m \s \pa^\m \s - 8 g^2 e^{2\vf} D_\m \r D^\m \r  - 4 P_{\m \un a} \bar P^{\m \un a} \nn \\
&\quad 
  - \fr14 e^{-\f} F_{\m \n} F^{\m \n}  - \fr18 \s \e^{\m\n\r\s} F_{\m \n} F_{\r \s}  - \fr14 e^{-\f} F^I_{\m \n} F_I^{\m \n}  - \fr18 \s \e^{\m\n\r\s} F^I_{\m \n} F_{\r \s I }  + 8  e^\f \tilde g^2 C^I C_I - 4 e^\f g^2 ( 1-e^\vf)^2  
\nn \\
&\quad
+ \bar \p_\m \g^{\m \n \r} D_{\n} \p_{\r} + \bar \l \g^\m D_\m \l  + \bar \l^I \g^\m D_\m \l_I +  \bar \c \g^\m D_\m \c +  \bar \z \g^\m D_\m \z \nn +  \bar \p{}_{\un a} \g^\m D_\m \p^{\un a}
 \nn \\
 &\quad
 - \fr1{2 \sq{2} } \big( \bar \p_\m \g^\n \g^\m \c + \bar \c \g^\m \g^\n \p_\m \big) \pa_\n \f 
- \fr1{2 \sq{2} } \big( \bar \p_\m \g^\n \g^\m \z + \bar \z \g^\m \g^\n \p_\m \big) \pa_\n \vf
\nn \\
&\quad
 +  \big( \bar \p_\m \g^\n \g^\m \p^{\un a} P_{\n \un a}  +  \bar \p_{\un a} \g^\m \g^\n \p_\m \bar P_\n{}^{\un a} \big) 
 \nn \\ 
 &\quad
+ \fr1{4} i e^{\f} \pa_{\m} \s \big ( \bar \p_\n \g^{\m \n \r} \p_\r + \sq 2 \bar \p_\n \g^{\m} \g^\n \c - \sq 2 \bar \c \g^\n \g^{\m} \p_\n  + 3 \bar \c \g^{\m} \c  -  \bar \z \g^{\m} \z +  \bar \l \g^{\m} \l  + \bar \l^I \g^{\m} \l_I +  \bar \p_{\un a} \g^{\m} \p^{\un a} \big)
 \nn \\
 &\quad
+ i g e^\vf D_\m \r \big ( \bar \p_\n \g^{\m \n \r} \p_\r + \sq 2 \bar \p_\n \g^\m \g^\n \z - \sq 2 \bar \z \g^\n \g^\m \p_\n  + 3 \bar \z \g^\m \z  -  \bar \c \g^\m \c - \bar \l \g^\m \l    - \bar \l^I \g^{\m} \l_I +  \bar \p_{\un a} \g^{\m} \p^{\un a} \big ) 
\nn \\
&\quad
+ \fr1{8 } e^{-\fr12 \f } F_{\m \n} \big( \sq 2 \bar \p_\r \g^{\m \n } \g^\r \l + \sq 2 \bar \l \g^\r \g^{\m \n} \p_\r + 2 \bar \c \g^{\m \n} \l - 2 \bar \l \g^{\m \n } \c  \big) 
\nn \\
&\quad 
+ \fr1{8 } e^{-\fr12 \f } F^I_{\m \n} \big( \sq 2 \bar \p_\r \g^{\m \n } \g^\r \l_I + \sq 2 \bar \l_I \g^\r \g^{\m \n} \p_\r + 2 \bar \c \g^{\m \n} \l_I - 2 \bar \l_I \g^{\m \n } \c  \big) 
\nn \\
&\quad 
- i g e^{\fr12 \f } ( e^\vf - 1) \big(\sq 2 \bar \p_\m \g^\m \l + \sq 2 \bar \l \g^\m \p_\m +  2 \bar \l \c - 2 \bar \c \l\big) 
- i  4  g e^{\fr12 \f } e^\vf ( \bar \l \z - \bar \z \l ) \nn \\ 
&\quad
-   \sq 2 \tilde g e^{  \fr12 \f} (   \bar \l^{I} { \p}{}^{\un a}   \x_{I \un a} + {\bar \p}{}_{\un a}  \l^{I} \bar \x_{I}{}^{ \un a } )   -    \tilde g e^{ \fr12 \f} C_{I} \big( \sq 2 \bar \l^I  \g^\m  \p_\m  +  \bar \p_\m  \g^\m \l^I  + 2  \bar \l^I \c  - 2 \bar \c \l^I  \big) \brs\ ,\label{redaction}
\ea
where
\ba
Q_\m &= \bar L^\un \a D_\m L_\un \a\ ,&
 Q_{ \m}{}^\un a{}_\un b &= - \bar L^{\un \a \un a} D_\m L_{\un \a \un b}\ ,\nn \\
P_{\m \un a} &= \bar L^\un \a D_\m L_{\un \a \un a}\ ,&
\bar P_{\m}{}^{ \un a} &=  L_\un \a D_\m \bar L^{\un \a \un a}\ ,\nn \\
D_\m L_{\un \a} &= \pa_\m L_{\un \a} - \tilde g A^I_\m T_{I \un \a}{}^\un \b L_{\un \b}\ ,&
D_\m \bar L^{\un \a} &= \pa_\m \bar L^{\un \a} +  \tilde g A^I_\m T_{I \un \b}{}^\un \a \bar L^{\un \b}\ ,\nn \\
D_\m L_{\un \a \un a} &= \pa_\m L_{\un \a \un a} - \tilde g A^I_\m T_{I \un \a}{}^\un \b L_{\un \b \un a}\ ,&
D_\m \bar L^{\un \a \un a} &= \pa_\m \bar L^{\un \a \un a} +  \tilde g A^I_\m T_{I \un \b}{}^\un \a \bar L^{\un \b \un a}\ ,\nn \\
\x^{I}{}_{\un a} &= \bar L^\un \a T^I{}_{\un \a}{}^\un \b L_{\un \b \un a}\ ,&
\bar \x^{I \un a} &= L_\un \a T^{I}{}_{ \un \b}{}^{\un \a} L^{\un \b \un b}\ , \nn \\
C^I &= \bar L^\un \a T^I{}_{\un \a}{}^{\un \b} L_{\un \b}\ .
\ea
The action \eqref{redaction} is supersymmetric under the transformations \eqref{4DHyperSusy} supplemented by
\ba
\d A^I_\m &= - \fr1{2 \sq{2} } e^{\fr12 \f} ( \bar \e \g_\m \l^I - \bar \l^I \g_\m \e)\ ,&
\d \l^I &= - \fr1{4 \sq{2} } e^{- \fr12 \f} F_{\a \b} \g^{\a \b} \e +   \sq 2 \tilde g e^{\fr12 \f} C^I \e\ .
\ea

\subsection{A Reduction Example}
\label{AReductionExample}

As an example of a reduction of this sort we consider starting with six-dimensional $N=1$ supergravity coupled to $28$ hypermultiplets, for which the scalars describe the coset $Sp(28,1)/(Sp(28) \otimes Sp(1))$, and one vector multiplet gauging the $U(1)_R$ R-symmetry as described in Section \ref{HypermultipletCouplings}.  The coset representatives then have a rigid left-acting $Sp(28,1)$ and local right-acting $Sp(28) \otimes Sp(1)$. The rigid left-acting $Sp(28,1)$ has an $Sp(28)$ subgroup and we chose to introduce 133 vector multiplets in order to gauge an $E_7$ subgroup of this $Sp(28)$ such that the irreducible $56$ of $Sp(28)$ remains an irreducible $56$ under $E_7$. This gives an action of the form \eqref{6DGaugedHyperAction}. We can then carry out the consistent reduction to four dimensions on a sphere and monopole background using the ansatz \eqref{GaugedHyperBosonicAnsatz} and \eqref{GaugedHyperFermionicAnsatz}.  

As described in Section \ref{SolutionConstraints}, it is only consistent to keep fluctuations in the hypermultiplet scalars that describe a coset $SU(28,1)/(SU(28) \otimes U(1))$ and, as described in \eqref{GaugeBreakingConditions}, the gauged $E_7$ symmetry is broken to the part that lies in $SU(28,1)$. This means that the surviving massless gauge symmetry will be the part of $E_7$ within $Sp(28)$ (such that the $56$ of $Sp(28)$ is the $56$ of $E_7$) that is also in the $SU(28)$ within $Sp(28)$ (such that the $56$ of $Sp(28)$ is the $28 + \bar {28}$ of $SU(28)$ ). 

To find the group describing this surviving gauge symmetry, we consider the $133$ $(56 \times 56)$ matrix generators of $E_7 $ within $Sp(28)$ whose construction is described in \cite{Cacciatori:2010dm}. We then restrict these generators to linear combinations such that the $(28 \times 28)$ off-diagonal blocks have only vanishing entries, so that \eqref{GaugeBreakingConditions} will be satisfied. This leaves $63$ linear combinations which we can split into $63$ $(56 \times 56)$ matrix generators $T^{I^\pr}$ of the surviving gauge group, which we normalise such that 
\ba
\text{tr}( T^{I^\pr}T^{J^\pr}) = \d^{I^\pr J^\pr} \ .
\label{NormGenerator1}
\ea
From these, we can calculate the structure constants of the surviving gauge group, which we find to satisfy 
\ba
f^{I^\pr K^\pr}{}_{L^\pr} f^{J^\pr L^\pr}{}_{K^\pr} = \fr43 \d^{I^\pr J^\pr} \ . 
\ea
This means that the matrix representation of the surviving group that we have obtained must have Dynkin index $\c$ and dual Coxeter number\footnote{For a set of group generators $T^a_{(r)}$ in a representation $r$ with a general normalisation, the Dynkin index $\c_r$ of $r$ may be defined as 
\ba
\text{tr} ( T^a_{(r)} T^b_{(r)} ) = - \chi_r g^{ab} \ , 
\ea
where $g^{ab}$ is the Cartan-Killing metric which is dependant on the normalisation of the generators $T^a_{(r)}$ in some particular representation.  Then, defining the structure constants by $[T_{(r)}^a , T_{(r)}^b ] = f^{a b}{}_c T^c_{(r)}$, the dual Coxeter number $\tilde h$ of the group is given by
\ba
f^{a c}{}_d f^{b d}{}_{c} &=  - \tilde h g^{a b} \ , 
\ea
\ie $\tilde h = \c_{\text{(adjoint)}}$. If we then chose to normalise the generators in the representation $r$ such that $\text{tr} ( T^a_{(r)} T^b_{(r)} ) = \d^{ab}$, as in \eqref{NormGenerator1}, then we find that 
\ba
f^{a c}{}_d f^{b d}{}_{c} &=  \fr{\tilde h}{\c_{r}} \d^{a b}  \ .
\ea
\vspace{-0.2cm}}
 ${\tilde h}$ such that $\frac{\tilde h}{\c} = \frac{4}{3}$. Considering all dimension $63$ subgroups of $E_7$ that have a dimension $56$ representation with this property reveals that the $63$ $(56 \times 56)$ matrix generators that we are left with must describe the $28 +\bar{28}$ of $SU(8)$, so $SU(8)$ becomes the surviving gauge group in this example. As indicated in Section \ref{HypermultipletCouplings}, the left-handed fermions arising from the reduction of the six-dimensional hypermultiplets carry an index in the fundamental representation of $SU(28)$, so the theory is chiral with respect to this symmetry. This is actually a situation where the left-handed charge conjugates of right-handed spinors that would have made the theory vectorlike are simply absent. In fact, the corresponding right-handed spinors were explicitly removed from the four-dimensional theory by the spinorial reduction ansatz \eqref{spinoransatz}.

An alternative starting six-dimensional theory which is of some interest is the $E_6 \otimes E_7 \otimes U(1)_R$ anomaly-free theory of Ref.\ \cite{RandjbarDaemi:1985wc}. This theory has $456$ hypermultiplets whose scalars describe the coset $Sp(456,1)/(Sp(456) \otimes Sp(1))$. The $E_7$ group here is the gauged part of the rigid $Sp(456)$ subgroup of the rigid left-acting $Sp(456,1)$ with the property that the irreducible 912 of $Sp(456)$ remains an irreducible 912 under $E_7$. The $U(1)_R$ of this theory is the gauged R-symmetry which will accommodate the monopole background in our reduction, while the $E_6$ remains external. The reduction of this theory on the sphere and monopole background using the ansatz \eqref{GaugedHyperBosonicAnsatz} and \eqref{GaugedHyperFermionicAnsatz} goes much like the reduction example considered above. This leads to a surviving massless gauge symmetry $G$ in four dimensions that is defined by the following conditions. It is product of the external $E_6$ times that part of the $E_7$ subgroup of $Sp(456)$ (where the $912$ of $Sp(456)$ remains an irreducible $912$ of $E_7$) that is also a subgroup of the $SU(456)$ within $Sp(456)$ (such that the irreducible $912$ of $Sp(456)$ becomes the $456 + \bar {456}$ of $SU(456)$). 

By analogy with the previous example, we propose that this reduction will give a reduced four-dimensional $N=1$ supergravity coupled coupled to chiral matter contained in multiplets whose scalars describe the coset $SU(456,1)/(SU(456) \otimes U(1)_R)$ of which the $SU(8)$ within the rigid $SU(456)$ is gauged by vector multiplets (such that the 456 of $SU(456)$ becomes the $420 + 36$ of $SU(8)$). The four-dimensional supergravity will also be coupled to additional external vector multiplets in the adjoint of $E_6$ together with a vector multiplet gauging a $U(1)_R$ whose gauge field becomes massive by the Stueckelberg mechanism as above. The left-handed fermions arising from the reduction of the six-dimensional hypermultiplets will carry an index in the fundamental of the gauged $SU(456)$ so the theory is clearly chiral with respect to this symmetry. 

\section{Conclusion}

In this paper, we have studied the varieties of consistent reductions available to a class of six-dimensional theories based on the original Salam-Sezgin model \cite{Salam:1984cj}. The key result is the bifurcation that we have found in Section \ref{Reduction} between the previous reduction which retains the sphere's $SU(2)$ gauge fields, but which proves to be non-chiral in four dimensions, and the new reduction which abandons the $SU(2)$ gauge fields in favour of a spontaneously broken $U(1)_R$, but allowing for a chiral theory in four dimensions. On general grounds, one might have expected \cite{Wetterich:1983ye,Witten:1983ux} some link between the reduction of a higher-dimensional theory on a background with nontrivial handedness and the preservation of chirality in the lower-dimensional reduced theory. However, such general considerations are not sufficient to elucidate details such as the loss of chirality when spherical isometry gauge fields are turned on. Only a detailed investigation of the reduction possibilities allows for the conditions of chirality preservation to be clearly established.

Although the new reduction carried out in Section \ref{Reduction} considers a consistent set of fluctuations about the very same background as that used previously in \cite{Gibbons:2003gp}, the result of the new reduction is very different. This is because our new ansatz considers fluctuations in the massive $U(1)_R$ gauge field (together with certain scalar modes) whereas the ansatz of \cite{Gibbons:2003gp} considered fluctuations in the massless $SU(2)$ gauge fields arising from the isometries of the sphere. A consistent ansatz turning on fluctuations in both these sectors simultaneously is not possible. This is because doing so would cause explicit functions of the internal space to appear in the reduced field equations. One place where this is clearly seen is in the reduction of the metric field equation. Here the ``miraculous'' cancellation described in \cite{Gibbons:2003gp} cannot occur once the $U(1)_R$ modes are turned on, thus dooming the consistency of the four-dimensional system with both $U(1)_R$ and $SU(2)$ gauge modes turned on. There is therefore a genuine bifurcation in the consistent reduction possibilities down to four dimensions.

The new consistent reduction scheme of Section \ref{Reduction} preserves four-dimensional chirality only in a fairly weak sense, because the chirality of the four-dimensional fermions is evidenced there only through couplings to the spontaneously broken, hence massive, $U(1)_R$ gauge field. Accordingly, in order in order to generate more clearly chiral couplings in the reduced theory, the class of six-dimensional models was extended in Section \ref{HypermultipletCouplings} by coupling in hypermultiplets and in Section \ref{GaugingHyperSymmetries} the gauging was extended by the inclusion of additional gauge multiplets. We then found that it is consistent to consider fluctuations in the a subset of the additional gauge fields, corresponding to a subgroup of the original gauge symmetries, together with the fluctuations of the massive $U(1)_R$. We note, however, that it still does not seem possible to turn on any fluctuations in the additional gauge symmetries if the $SU(2)$ fluctuations arising from the sphere's isometries are turned on instead of the $U(1)_R$ fluctuations. This reduction inconsistency happens in an exactly analogous way to that seen when trying to simultaneously turn on the $SU(2)$ and $U(1)_R$  gauge fluctuations. For this reason, the new branch of consistent fluctuations is the more potentially interesting for four-dimensional phenomenology.

\section*{Acknowledgments}

K.S.S. would like to thank the Niels Bohr Institute and the Laboratoire de Physique Th\'eorique de l'\'Ecole Normale Sup\'erieure for hospitality during the course of the work. The research of C.N.P. was supported in part by DOE grant DE-FG03-95ER40917. The research of K.S.S. was supported in part by the STFC under rolling grant ST/G000743/1.

\begin{appendix}
\section{Conventions}
In this work, we have used conventions where 
\ba
* \o_{(r)} &= \fr1{r! (d-r)! } \e_{\a_1 \ldots \a_{ d-r }}{}^{\b{1} \ldots \b_r} \o_{\b_1 \ldots \b_r } e^{\a_1} \we \ldots \we e^{\a_{ d-r} } 
\ea
\ba
\o_{(r)} \we \h_{(s)} &= \fr{1}{r! s!} \o_{\a_1 \ldots \a_r} \h_{\b_1 \ldots \b_s} e^{\a_1} \we \ldots \we e^{\a_r} \we e^{\b_1} \we \ldots e^{\b_s} 
\ea
\ba
\wh R_{MN} &= \wh R^R{}_{M R N}  & R_{MN}{}^{AB} &= \pa_{M} \o_N{}^{AB} + \ldots 
\ea
\ba  
\wh \G_7 &=\wh  \G_0 \ldots \wh  \G_5\ , & \wh {\bar \c} &= i \wh \c^{\dagger} \wh \G_0
\ea 
\ba
\g_5 &= i \g_0 \ldots \g_3 & \bar \c &= i \c^\dagger \g_0  
\ea
\ba
\wh {\bar \c}^\un A &= i  (\wh \c_\un A ) ^{\dagger} \wh \G_0  = ( \wh { \c}^\un A)^T & \wh {\bar \p}^{\uh a} &= i ( \wh \p_{\uh a}  )^{\dagger} \wh \G_0 = ( \wh {\bar \p}^{\uh a})^T  & 
\bar \p_{\un a} &= i ( \p^{\un a}  )^{\dagger} \wh \G_0 
\ea
\ba
 \g^{\a\b\g\d} &= i \g_5 \e^{\a\b\g\d} &\wh \e^{012345} &= \e^{0123} = 1 
 \ea
 \ba
[\G_A, \G_B] &= 2\h_{AB}& \h_{AB} &= \text{diag} ( - , +, ..., +) 
\ea
\ba
\wh \G_\a &= \g_\a \ct \s_3 &  \wh \G_a &= \id \ct \s_a &
\wh \G_7 &= \g_5 \ct \s_3 
\ea
\ba
\wh C \wh \G_A \wh C^{-1} &= - \wh \G_A^T & \wh C &= C \ct \s_2\ .
\ea

\end{appendix}


\end{document}